# Spin canting in nonlinear terahertz magnon dynamics revealed by magnetorefractive probing in orthoferrite

*† **Takayuki Kurihara**[1], †Motoaki Bamba[2,3], Hiroshi Watanabe[1,4], Makoto Nakajima[1,5] and Tohru Suemoto[1,6]

[1]*Institute for Solid State Physics, The University of Tokyo, 5-1-5 Kashiwanoha, Kashiwa, Chiba 277-8581, Japan*
[2]*Department of Physics I, Kyoto University, Kitashirakawa Oiwake-cho, Sakyo-ku, Kyoto 606-8502, Japan*
[3]*The Hakubi Center for Advanced Research, Kyoto University, Yoshida-honmachi, Sakyo-ku, Kyoto 606-8501, Japan*
[4]*Graduate School of Frontier Biosciences, Osaka University, 1-3 Yamadaoka, Suita, Osaka 565-0871, Japan*
[5]*Institute of Laser Engineering, Osaka University, 2-6 Yamadaoka, Suita, Osaka 565-0871, Japan*
[6]*Department of Engineering Science, The University of Electro-Communications, Chofu, Tokyo 182-8585, Japan*
* Corresponding author T.K. (takayuki.kurihara@issp.u-tokyo.ac.jp)
† These authors contributed equally for this work

**We excite the spin precession in rare-earth orthoferrite $YFeO_3$ by the magnetic field of intense terahertz pulse and probe its dynamics by transient absorption change in the near infrared. The observed waveforms contain quasi-ferromagnetic-mode magnon oscillation and its second harmonics with a comparably strong amplitude. The result can be explained by dielectric function derived from magnetorefractive Hamiltonian. We reveal that the strong second harmonic signal microscopically originates from novel dynamics of the quasi-ferromagnetic mode magnon at nonlinear regime, wherein spin canting angle periodically oscillates.**

Control of spin dynamics at ultrafast time scales has been one of the central topics in modern magnetism research. Excitation by femtosecond laser pulses drastically changes the intrinsic magnetic properties such as net magnetization, exchange and anisotropy [1]–[9]. Terahertz (THz) pulses offer unique advantage that their magnetic fields can excite spin dynamics resonantly at femto- to picosecond time scales while avoiding unwanted electronic and lattice heating due to its low photon energy [10]–[16]. Since the advancement of the intense THz light sources in the last decade, coherent THz pulses with peak electric- and magnetic fields in the order of MV/cm and ~1 Tesla has become available [17]–[24]. Such intense magnetic field pulses can be used to study the nonlinear dynamics of spin systems in solids. For example, recent works report macroscopic control of magnetization [12], [13], [25], magnon softening [26] and magnon-induced terahertz second harmonic generation [27] to name a few.

The detection of magnetization dynamics $\boldsymbol{M}(t)$ induced by light pulses commonly relies on magnetooptical Faraday- and Kerr effects [28]–[30] or magnetic dichroism [31]. These effects are based on the off-diagonal components of the complex dielectric permittivity tensor, which have proportional or odd-order dependences with respect to the induced magnetization change ($\boldsymbol{M}^{2N-1}$, $N$ = integer) as derived by Onsager [32]–[34] from the arguments based on time-reversal symmetry and energy conservation. However, since the spin deflection angles excited even by the state-of-the-art table-top THz sources are small in most cases [13], [35], in this detection scheme the possible nonlinear responses that scale super-linearly with the incident THz fields are potentially hindered when linear spin dynamics dominates the signal [35]. From this viewpoint, a probing method to sensitively extract the nonlinear contribution of the spin dynamics is of practical importance.

On the other hand, the diagonal components of the permittivity tensor are of even order ($\boldsymbol{M}^{2N}$) as derived by Onsager [32]–[34]. This term gives rise to the so-called quadratic magnetorefractive effect, wherein the induced magnetization causes optical birefringence or absorption [4], [36], [37]. Because the linear component ($\boldsymbol{M}^{1}$) does not directly appear in the diagonal components of the complex permittivity, it is expected that probing the absorption change allows us a precise investigation of the nonlinear spin dynamics with quadratic field-dependence. While magnetorefractive effect has been previously employed for the observation of the photoinduced modification of the exchange parameters in correlated spin systems [4], [36], its application to the observation of nonlinear spin dynamics excited by resonant THz magnetic fields has not been achieved to the best of our knowledge.

Here, we study the THz-induced nonlinear magnetorefractive effects by probing the absorption change. We excite the quasi-ferromagnetic magnon mode in rare-earth orthoferrite $YFeO_3$ by the intense THz magnetic fields and detect its dynamics by the absorption change of a near-infrared probe pulse. As a result, the transient absorption signal shows a strong second harmonic oscillation of the originally excited magnon mode at comparable amplitude with the fundamental mode. Through both the numerical simulation and analytical calculation, we show that the second harmonics signal can be explained by taking into account the phenomenological Hamiltonian representing magnetorefractive interaction between the sublattice spins and the probe electromagnetic fields. It indicates a new type of nonlinear motion of the quasi-ferromagnetic magnon mode where the canting angle between sublattice spins periodically changes at twice the precession frequency, possibly suggesting its nonlinear mixing with the other magnon mode, the quasi-antiferromagnetic oscillation.

We study a float-zone-grown single crystal of the rare-earth orthoferrite $YFeO_3$. Orthoferrite is a typical weak antiferromagnet, in which antiferromagnetically ordered sublattice spins are slightly canted due to Dzyaloshinskii-Moriya interaction and result in a net magnetization $\boldsymbol{M}$ [38]–[41]. At room temperature, the spin configuration forms the so called $\Gamma_4$ phase, wherein sublattice spins align parallel to the $a$-axis and the magnetization $\boldsymbol{M}$ along $c$-axis [FIG. 1 (a)]. Orthoferrites possess two isolated magnon modes in the sub-THz frequency region. One is the quasi-ferromagnetic mode (FM), which corresponds to precessional motion of the total magnetization $\boldsymbol{M}$. The other one is called the quasi-antiferromagnetic mode (AFM), which can be viewed as the oscillation of the vector length of $\boldsymbol{M}$. The FM and AFM have different selection rules in terms of polarization. The FM is excited by magnetic field when $\boldsymbol{B}_{THz} \perp$ M, that is $\boldsymbol{B}_{THz}$ // $a$ or $b$ at room temperature. The AFM is excited by $\boldsymbol{B}_{THz}$ // M, that is $\boldsymbol{B}_{THz}$ // $c$. The resonance frequencies of the F- and AF modes in $YFeO_3$ are $f_{FM}$ = 0.3 THz and $f_{AFM}$ = 0.52 THz at room temperature [10] . The sample is *c*-plane cut and have the thickness of approximately 100 μm.

Our experimental setup is schematically illustrated in FIG. 1 (a). The intense THz pulses are generated by optical rectification in $LiNbO_3$ crystal using wave front tilting technique [19]. It is pumped by the output from Ti:Sapphire regenerative amplifier (*Legend Elite* from Coherent inc.), which has central wavelength at 800 nm, pulse width of approximately 100 fs, repetition rate at 1 kHz, and pulse energy of 5 mJ. The resulted THz transient has a single-cycle waveform and a spectrum covering 0 - 1.8 THz, as shown in FIG. 1 (b, c). The terahertz magnetic field is linearly polarized parallel to $\boldsymbol{B}_{\mathrm{THz}}$ // *b*, and the amplitude of the THz pulse is varied by two wire grid polarizers. Part of the regenerative amplifier output are mechanically delayed and used as probe pulses. In order to cancel out the intensity fluctuation of the light sources and obtain sufficient signal to noise ratio, we employ balanced photodetection by splitting the probe pulse in two paths and measuring the difference between the pulses transmitted through the sample and the reference which is directly sent into the other photodiode. The probe pulse $\boldsymbol{E}_P$ is linearly polarized parallel to the crystal *a*-axis, unless otherwise stated. All measurements are performed at room temperature. No external magnetic fields are applied on the samples.

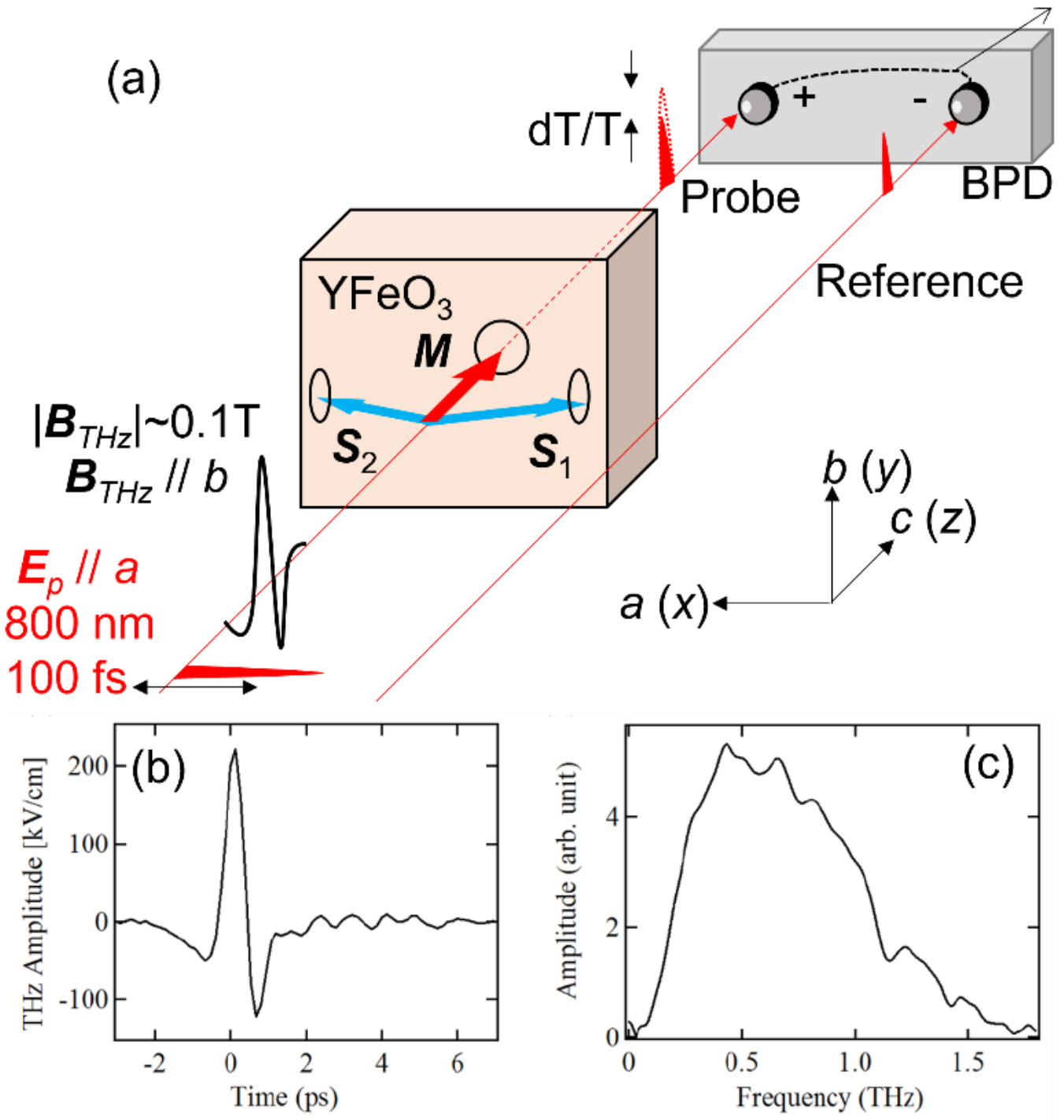

Figure 1: (a) Experimental setup of the intense THz pump–induced magnetorefraction measurement. The blue and red arrows indicate the sublattice magnetizations $\boldsymbol{S}_{1,2}$ and net magnetization $\boldsymbol{M}$, respectively. RT: Room temperature, BPD: Balanced photodetector. (b) Typical waveform of the THz electric field measured with electrooptic sampling in a GaP (110) single crystal with 400 μm thickness. (c) Fourier spectrum of (b).

Figure 2 (a) shows the transient absorption waveforms measured in $YFeO_3$ at incident electric field amplitudes of 110 kV/cm, 165 kV/cm, and 220 kV/cm. The THz pulses are incident at $t$ = 0 ps, which is marked by a sharp peak. This signature increases quadratically with the incident field, and is ascribed to a third-order optical nonlinearity due to the incident THz electric field [43]. In the region $t$ > 0 ps, an oscillation at a period of 3.3 ps is observed. This period matches that of the FM magnon at $f_{FM}$ = 0.3 THz and indicate that this oscillation originates from the precession of the total magnetization $\boldsymbol{M}$ around $c$-axis. The envelope function shows increase after pumping and does not decay exponentially as contrasted to normal spin dynamics. This characteristic feature is commonly observed in the spin dynamics of orthoferrites under THz excitation. It is ascribed to the electromagnetic interaction between propagating magnon modes and the incident THz magnetic field. Detailed investigation of this feature is provided in previous works such as [44], [45], and is therefore not the main scope of the present work.

The waveform is strongly dependent with the incident field strength. Under the strong THz excitation amplitudes, the waveform becomes highly asymmetric. In this regime, there appears an additional oscillatory signature that is synchronized with the original FM but is half the period. As can be seen in the Fourier spectra in FIG. 2 (b), this oscillation has a frequency at 0.6 THz, which precisely matches the second harmonic (SH) of $f_{FM}$ = 0.3 THz. It also clearly differs from the other magnon mode AFM, which sets in at 0.53 THz [10], indicating that it originates from the FM dynamics. The field dependence of the spectral peaks at 0.3 THz and 0.6 THz are plotted in FIG. 1 (c). The 0.3 THz peak is linearly dependent with the incident THz field, while the 0.6 THz peak scales quadratically. It should also be noted that the appearance of the second-harmonic signal is also confirmed in a different orthoferrite $ErFeO_3$ (Supplemental Material A). It suggests that the underlying mechanism of this SH phenomenon is general to the orthoferrites, regardless of the type of the constituent rare-earth ions. Both the fundamental and the second harmonic oscillations are observed only when the probe pulse is polarized along $\boldsymbol{E_p}$ // $a$ (Supplemental Material B).

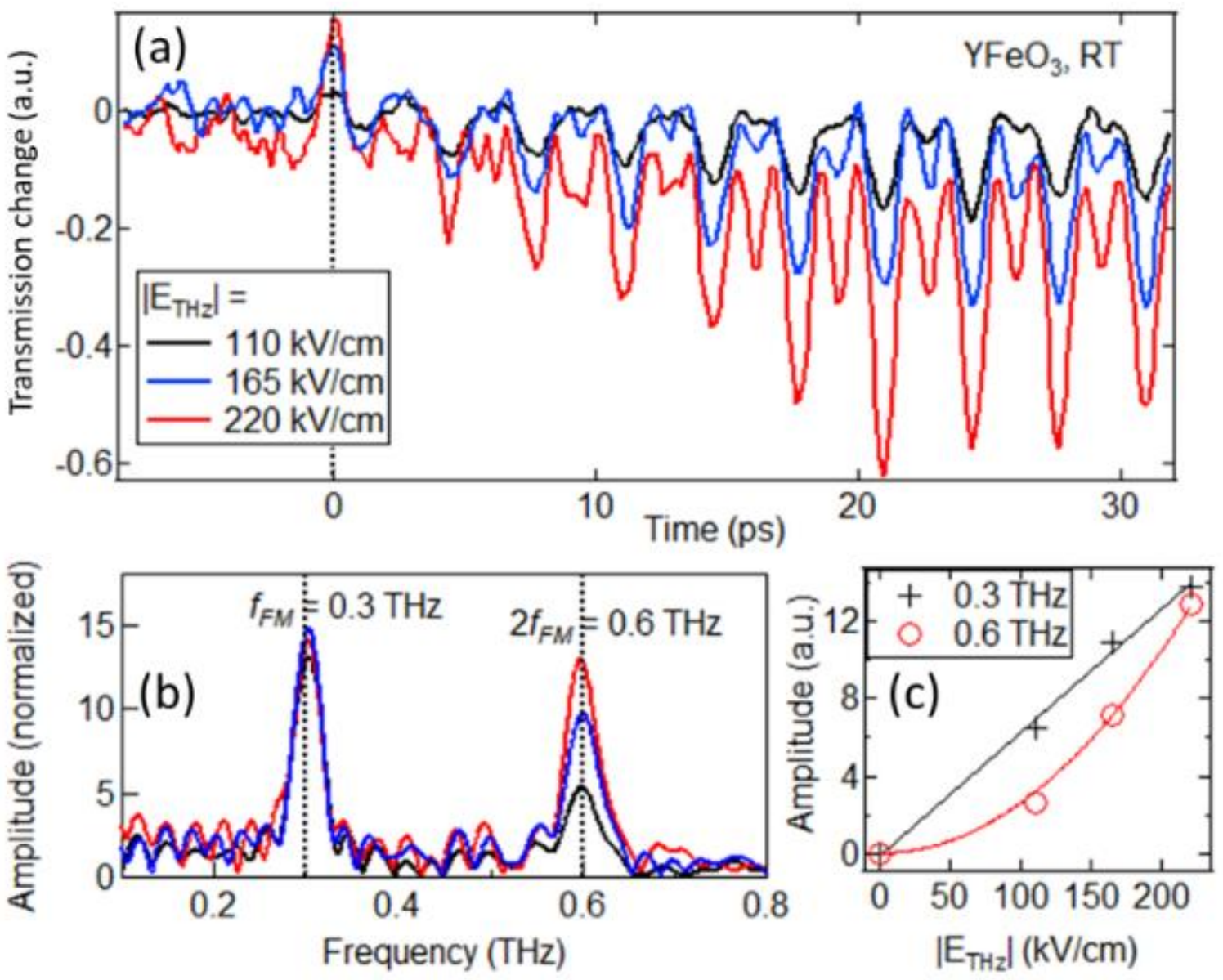

Figure 2: (a) Transient absorption signals induced by THz excitation in $YFeO_3$ sample under various THz field strengths. (b) Fourier spectrum of (a) normalized to the peak at 0.3 THz. (c) Spectral amplitudes of 0.3 THz and 0.6 THz peaks as functions of incident THz field strength. Fitting curves with linear (0.3 THz) and quadratic (0.6 THz) dependence are also plotted.

In order to understand the mechanism of the observed second harmonics signal in detail, we perform numerical calculations of the spin dynamics based on the two-sublattice model which is commonly used to describe orthoferrites [39], [40]. Substituting the calculated spin dynamics into the time-dependent dielectric permittivity function derived from magnetorefractive interaction between the spins and the probe electric field, we reproduce the experimentally measured second-harmonics signals.

First, we consider the Hamiltonian of the $Fe^{3+}$ spins as:

$$\mathcal{H}_{\mathrm{Fe}} = zJ_{\mathrm{Fe}}\boldsymbol{S}_1 \cdot \boldsymbol{S}_2 - zD_{\mathrm{Fe}}(S_{1x}S_{2z} - S_{1z}S_{2x}) - A_x\left({S_{1x}}^2 + {S_{2x}}^2\right) - A_z\left({S_{1z}}^2 + {S_{2z}}^2\right) \\ -A_{xz}(S_{1x}S_{1z} - S_{2x}S_{2z}), \quad (1)$$

where $z$ = 6 is the number of nearest neighboring $Fe^{3+}$ sites, $J_{\mathrm{Fe}}$ = 4.96 meV and $D_{\mathrm{Fe}}$ = 0.107 meV are isotropic and antisymmetric exchange interaction strengths, respectively, and $A_{x;z;xz}$ are the magnetic anisotropy energies ($A_x$ = 0.00288 meV, $A_z$ = 0.0008 meV, $A_{xz}$ = 0 meV). Each parameter is adjusted from the literature values ([13], [46]–[48]) to reproduce the FM and AFM resonance frequencies of $YFeO_3$ at room temperature. The equation of motion of $\boldsymbol{S}_{j=1,2}$ for the Hamiltonian including external magnetic flux $\boldsymbol{B}_{\mathrm{ext}}(t)$ is expressed as:

$$\gamma^{-1}\dot{\boldsymbol{S}}_j = -\boldsymbol{S}_j \times \left[\boldsymbol{B}_{\mathrm{ext}}(t) + \boldsymbol{\nabla}_{\boldsymbol{S}_j}\mathcal{H}_{\mathrm{Fe}}/(g\mu_{\mathrm{B}})\right], \quad (2)$$

where $g$ is the $g$-factor, $\mu_{\mathrm{B}}$ is the Bohr magneton, and $\gamma = g\mu_{\mathrm{B}}/\hbar$ is the gyromagnetic ratio.

In order to describe the quadratic magnetorefractive effect observed experimentally, we consider the following form of Hamiltonian representing nonlinear interaction between the spin dynamics and the electric field of probe light [4], [36]:

$$\mathcal{H}' = \frac{1}{2}\sum_{\xi,\xi'=x,y,z} E_{\xi}(t)E_{\xi'}(t)\{\alpha_{\xi,\xi'}\boldsymbol{S}_1(t)\cdot\boldsymbol{S}_2(t) + \boldsymbol{\beta}_{\xi,\xi'}\cdot[\boldsymbol{S}_1(t)\times\boldsymbol{S}_2(t)]\}. \quad (3)$$

Here, $E_\xi$ is the electric field in the $\xi = \{x, y, z\}$ direction. $\alpha_{\xi,\xi'}$ and $\beta_{\xi,\xi'}$ are coefficients for the isotropic and antisymmetric exchange interactions that connect the electric field $E_\xi$ and the spins $\boldsymbol{S}_{1,2}$. This Hamiltonian phenomenologically originates from the virtual transition of $O^{2-}$–$Fe^{3+}$ charge transfer [4] (Supplemental Material C). We can rewrite Eq. (3) in the form of dielectric permittivity tensor as:

$$\mathcal{H}' = \sum_{\xi,\xi'} \varepsilon_{\xi,\xi'}(t)E_\xi(t)E_{\xi'}(t)/2. \quad (4)$$

From Eqs. (3) and (4), the change of permittivity induced by the spin dynamics can be deduced as

$$\varepsilon_{\xi,\xi'}(t) \equiv \alpha_{\xi,\xi'}\boldsymbol{S}_1(t)\cdot\boldsymbol{S}_2(t) + \boldsymbol{\beta}_{\xi,\xi'}\cdot[\boldsymbol{S}_1(t)\times\boldsymbol{S}_2(t)]. \quad (5)$$

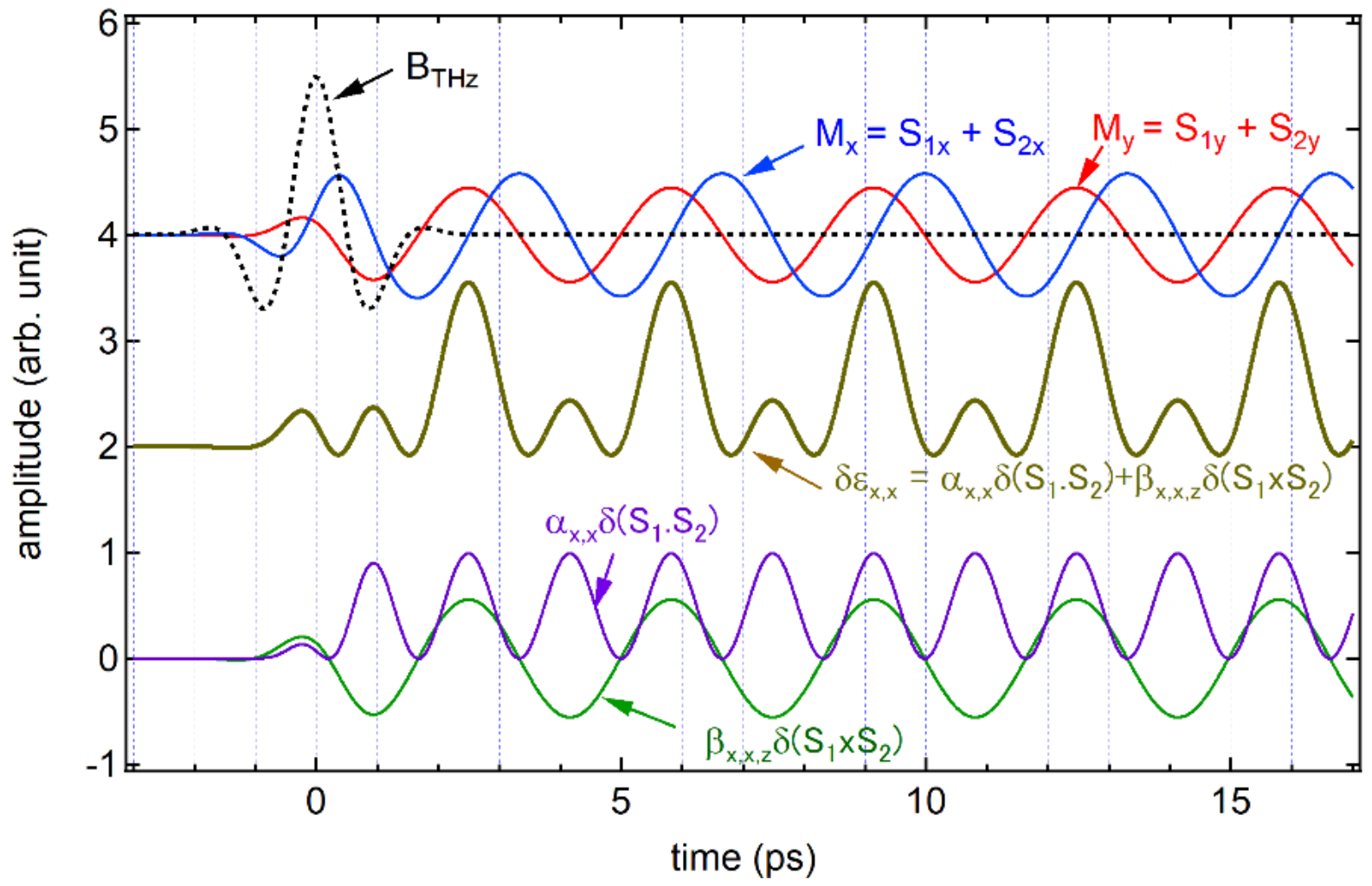

Figure 3: Numerically calculated waveforms of THz magnetic field $B_{\text{THz}}$ (black dashed curve), macroscopic magnetization $M_x$ (blue) and $M_y$ (red), dielectric function $\varepsilon_{x,x}$ (dark yellow) and its constituent terms, dot- (violet) and cross product terms (green). Each curve is offset for clarity.

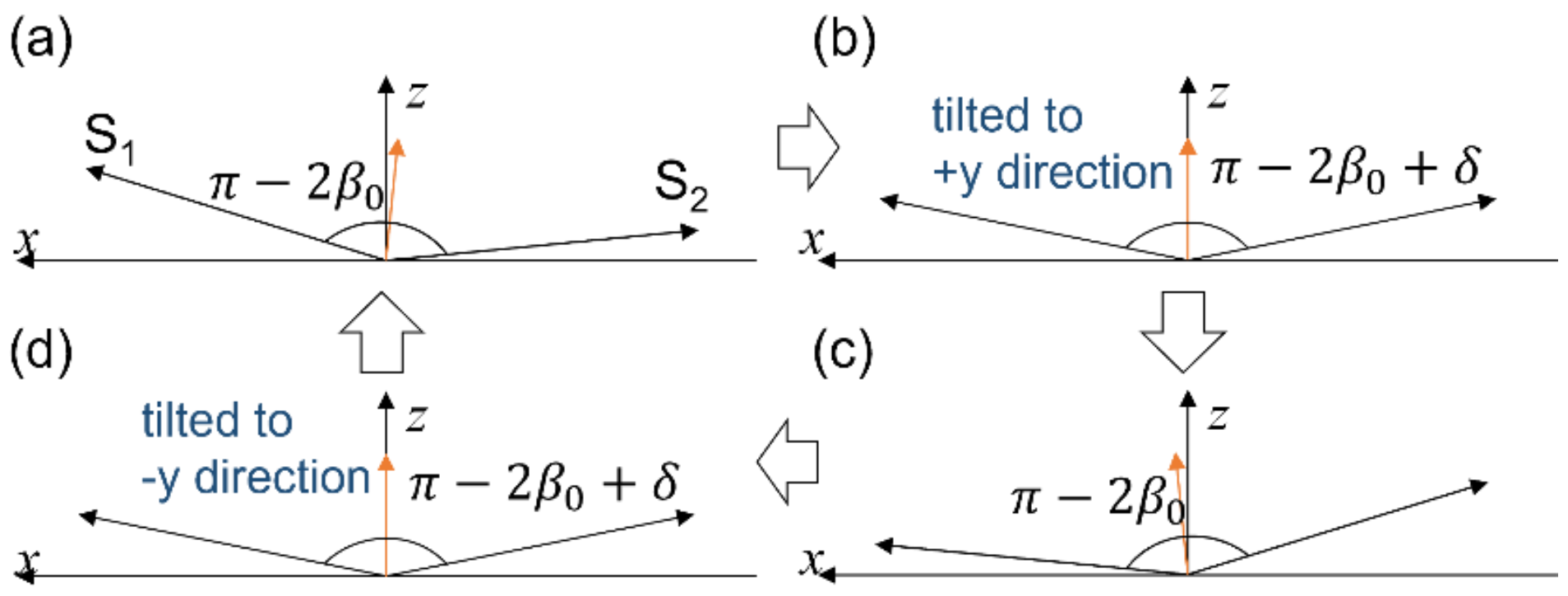

Figure 4: Schematic representation of the intra-angle modulation of the spin canting angle during the FM precession, corresponding to $t$ = 1.6 ps (a), $t$ = 2.5 ps (b), $t$ = 3.3 ps (c), and $t$ = 4.1 ps (d) in Figure 3.

The spin dynamics numerically calculated from Eqs. (1) and (2) are plotted in FIG. 3. The blue and red curves $M_x$ and $M_y$ correspond to the *x*- and *y*-components of the macroscopic magnetization $\boldsymbol{M} = \boldsymbol{S}_1 + \boldsymbol{S}_2$, respectively, excited by a THz magnetic field pulse (black dotted waveform $B_{\mathrm{THz}}$) with a central frequency of 0.5 THz and a peak amplitude of 0.1 T, mimicking the experimental waveform. Calculated $M_x$ and $M_y$ show clear sinusoidal oscillations at the fundamental frequency $f_{FM}$ = 0.30 THz with their phase shifted by 90 degrees from each other, indicating that the dynamics of the individual spins $\boldsymbol{S}_{1,2}$ is dominated by a linear response of the FM precession. Substituting the calculated spin dynamics into Eq. (4), as shown by the yellow curve in FIG. 3, we obtain the transient absorption change $\delta\varepsilon_{x,x} \equiv \mathrm{Im}\left[\varepsilon_{x,x}(t) - \bar{\varepsilon}_{x,x}\right]$, where $\bar{\varepsilon}_{x,x}$ is the equilibrium permittivity without the THz pulse. Since in our experiment only the $a$-polarized probe was found sensitive to the spin dynamics [Supplemental Material B], we take $\mathrm{Im}\left[\alpha_{x,x}\right]$ and $\mathrm{Im}\left[\beta_{x,x,z}\right]$ as the most relevant components in the permittivity tensor that are observable. The ratio of $\mathrm{Im}\left[\alpha_{x,x}\right]$ and $\mathrm{Im}\left[\beta_{x,x,z}\right]$ was tuned at $|\mathrm{Im}\left[\alpha_{x,x}\right]/\mathrm{Im}\left[\beta_{x,x,z}\right]| = 2 \times 10^4$ to reproduce the experimentally observed amplitude ratio between the fundamental and second harmonic signals at maximum incident THz field strength in FIG. 2 (c). In contrast to the original spin dynamics $M_x$ and $M_y$, the induced permittivity change exhibits a clear asymmetric waveform, reproducing the experimental observation. In this way, the quadratic response is well reproduced by the nonlinear magnetorefractive interaction introduced in Eq. (3).

In order to investigate the clearly asymmetric permittivity change in further detail, we break down the yellow curve in FIG. 3 into the separate constituent terms in Eq. (4), i.e., the isotropic term $\delta\left(\boldsymbol{S}_1(t) \cdot \boldsymbol{S}_2(t)\right) = \overline{\boldsymbol{S}}_1 \cdot \delta\boldsymbol{S}_2(t) + \delta\boldsymbol{S}_1(t) \cdot \overline{\boldsymbol{S}}_2 + \delta\boldsymbol{S}_1(t) \cdot \delta\boldsymbol{S}_2(t)$ and the antisymmetric one $\delta\left(\boldsymbol{S}_1(t) \times \boldsymbol{S}_2(t)\right) = \overline{\boldsymbol{S}}_1 \times \delta\boldsymbol{S}_2(t) + \delta\boldsymbol{S}_1(t) \times \overline{\boldsymbol{S}}_2 + \delta\boldsymbol{S}_1(t) \times \delta\boldsymbol{S}_2(t)$. $\overline{\boldsymbol{S}}_{1,2}$ and $\delta\boldsymbol{S}_{1,2}(t)$ represent the equilibrium- and dynamical-components of the sublattice spins, respectively. The results are plotted as violet and green curves in FIG. 3. For the antisymmetric term, we plot only the *z*-component which is the most dominant among all three spatial axes by orders of magnitude. FIG. 3 shows that the antisymmetric term $\delta\left(\boldsymbol{S}_1(t) \times \boldsymbol{S}_2(t)\right)$ contains mostly the linear ($\omega$) component. On the other hand, the isotropic term $\delta\left(\boldsymbol{S}_1(t) \cdot \boldsymbol{S}_2(t)\right)$ is dominated by the second harmonic ($2\omega$) component. These results can be understood by the fact that the equilibrium and dynamical components [$\overline{\boldsymbol{S}}_{1,2}$ and $\delta\boldsymbol{S}_{1,2}(t)$] are perpendicular to each other during the FM precession.

The presence of a finite amplitude of the isotropic term $\delta\left(\boldsymbol{S}_1(t) \cdot \boldsymbol{S}_2(t)\right)$ gives us an interesting insight into the nonlinear spin dynamics and magnon–magnon interactions. In the limit of linear response, the spins' canting angle $\beta_0$ remains fixed during the FM precession in orthoferrites [39], and hence $\delta\left(\boldsymbol{S}_1(t) \cdot \boldsymbol{S}_2(t)\right) = 0$. In contrast to such conventional picture of the FM trajectory, our result clearly shows that the relative angle between the sublattice spins $\boldsymbol{S}_1$ and $\boldsymbol{S}_2$ periodically oscillates at twice the precession frequency during the FM dynamics, when the precession amplitude is large. Taking this effect into account, a modified dynamics of FM derived from the calculated waveforms $M_x$, $M_y$ and $\varepsilon_{xx}$ is schematically illustrated in FIG. 4. When both spins $\boldsymbol{S}_1$ and $\boldsymbol{S}_2$ are in the *x*-*z* plane (FIG. 4 (a) and (c)), the canting angle is $\pi - 2\beta_0$, the same as its equilibrium value. As the precession evolves and the total magnetization directs towards either +*y* or -*y* axis, the canting angle slightly increases (FIG. 4 (b) and (d)). Because the spins tilt along the *y* axis twice during a cycle, the canting angle oscillates with double the precession frequency, leading to the observed second harmonics through magnetorefractive effect.

From another viewpoint, the above-revealed periodic canting in the FM can be understood as a nonlinear mixing of the FM magnon with the other magnon mode, AFM. In the AFM, a periodic change of the canting angle occurs even in the limit of linear response. Thus, the fundamental of AFM and second harmonic of the FM partially shares the same geometrical motion, which should lead to the nonlinear mixing of the two modes as the FM is strongly excited. In fact, analytical calculation up to second order perturbation shows that there exists a driving force term in the AFM equation of motion which scales quadratically with the FM amplitude, as theoretically investigated in detail in the Supplemental Material D-F.

To conclude, we have excited the FM magnon in the orthoferrites $YFeO_3$ and $ErFeO_3$ and measured the transient absorption change in the near infrared. The FM magnon dynamics and its second harmonic have been observed with comparable amplitudes. The second harmonic dynamics are qualitatively reproduced by taking into account the magnetorefractive interaction between the sublattice spins and the probe electromagnetic fields. From quantitative numerical calculation, it has been shown that the sublattice spin angle, which has been commonly assumed fixed during the FM precession, oscillates at its second harmonic frequency, suggesting possible mixing of FM and AFM at large precession amplitude. The scheme of THz pump–magnetorefractive probing demonstrated here paves way for selectively detecting the coherently excited nonlinear spin dynamics and magnetooptical effects in various correlated spin systems in the future.

## Acknowledgements

This work has been funded by the Japan Society for the Promotion of Science (JSPS) KAKENHI (JP20K22478, JP21K14550 and JP20H02206). The single crystals of orthoferrite used in the study have been grown using facilities in Materials Synthesis Section at ISSP, The University of Tokyo. The authors thank Prof. H. Mino at Chiba Univ. for the insightful discussions.

## Author Contributions

T.K. conceived the experimental idea, performed measurement and wrote manuscript. M.B. performed theoretical calculation and wrote manuscript together with T.K. H.W and M.N. constructed the THz generation setup and have grown the orthoferrite samples used in the study. T.S. supervised the project together with T.K.

# Supplemental Material: Spin canting in nonlinear terahertz magnon dynamics revealed by magnetorefractive probing in orthoferrite

Takayuki Kurihara,[1] Motoaki Bamba,[2,3] Hiroshi Watanabe,[1,4] Makoto Nakajima,[1,5] and Tohru Suemoto[1,6]

[1]*Institute for Solid State Physics, The University of Tokyo, 5-1-5 Kashiwanoha, Kashiwa, Chiba 277-8581, Japan*
[2]*Department of Physics I, Kyoto University, Kitashirakawa Oiwake-cho, Sakyo-ku, Kyoto 606-8502, Japan*
[3]*The Hakubi Center for Advanced Research, Kyoto University, Yoshida-honmachi, Sakyo-ku, Kyoto 606-8501, Japan*
[4]*Graduate School of Frontier Biosciences, Osaka University, 1-3 Yamadaoka, Suita, Osaka 565-0871, Japan*
[5]*Institute of Laser Engineering, Osaka University, 2-6 Yamadaoka, Suita, Osaka 565-0871, Japan*
[6]*Department of Engineering Science, The University of Electro-Communications, Chofu, Tokyo 182-8585, Japan*

This Supplemental Material includes six Appendices. In Appendix A, the measurement results for $ErFeO_3$ sample is presented. In Appendix B, dependence of the measured signals on the THz and probe polarizations are discussed. In Appendix C, we will explain the physical meaning of the magnetorefractive Hamiltonian used in the main text. In Appendix D, the magnon modes in rare-earth orthoferrites excited by terahertz (THz) pumping will be discussed. In Appendix E, how the magnons modifies the permittivity through the magnetorefractive Hamiltonian will be discussed. In Appendix F, we will see one magnon in the quasi-antiferromagnetic mode (AFM) can be excited by two magnons in the quasi-ferromagnetic mode (FM).

## Appendix A: Second harmonic oscillation in $ErFeO_3$

The appearance of the second-harmonic signal is also found in a $ErFeO_3$ sample. Here, the waveforms are measured under the same polarization configurations for the incident THz and probe pulses as in the case of $YFeO_3$ measurement. The $ErFeO_3$ sample is c-plane cut and has a thickness of 100 $\mu$m. After t = 0 ps, an oscillation with frequency matching that of the FM magnon in $ErFeO_3$ $f_{\mathrm{FM}}^{\mathrm{ErFeO_3}} = 0.38$ THz is observed [1]. At high THz field strengths, (see e.g., 440 kV/cm in Fig. 1 (a)) the asymmetric distortion of the periodic pattern due to the appearance of SH is clearly reproduced. Even though the SH signal is weaker than the case of $YFeO_3$, the corresponding normalized spectrum clearly exhibits SH peak at $2f_{\mathrm{FM}}^{\mathrm{ErFeO_3}} = 0.76$ THz [Fig. 1 (b)]. This observation suggests that the underlying mechanism of this SH phenomenon is general to the orthoferrites, regardless of the type of the constituent rare-earth ions.

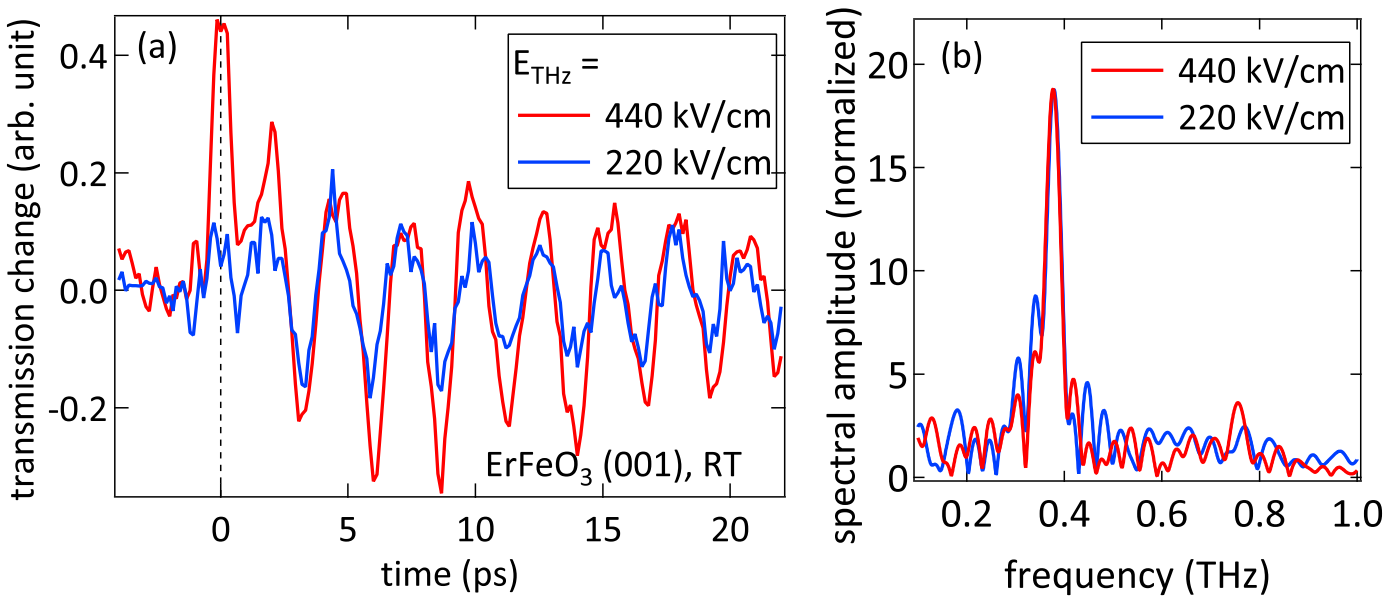

FIG. 1. (a) THz-induced transient absorption at 800 nm measured in $ErFeO_3$ crystal. (b) Fourier spectra of (a), normalized at fundamental peak frequency of 0.38 THz.

## Appendix B: Polarization dependence of the signal

The polarization dependence of the observed signals is plotted in FIG. 2. When the polarization of the incident THz pulse $E_{\mathrm{THz}}$ is rotated from $E_{\mathrm{THz}}$ // $a$ to $E_{\mathrm{THz}}$ // $b$ , the phase of the FM oscillation shifts by 90 degrees, as expected for the precession motion of the macroscopic magnetization. Additionally, it is found that both the fundamental FM and the second harmonic signatures are sensitive to the polarization $E_p$ of the probe pulse. While the signals are clearly detectable for $E_p \,//\, a$, it disappears in the case of $E_p \,//\, b$. This selection rule indicates that the sensitivity of the dielectric tensor components along each axis on the magnetization change is different, probably due to crystal symmetry.

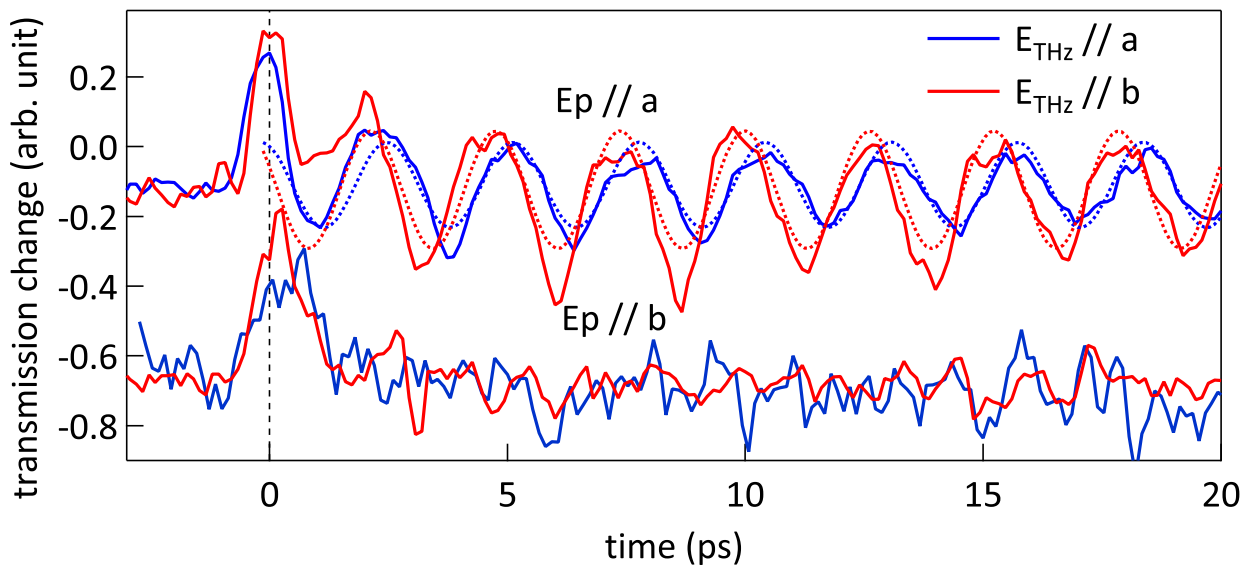

FIG. 2. Transient absorption waveforms measured in $ErFeO_3$ for $E_p \,//\, a$ (upper waveforms) and $E_p \,//\, b$ (lower waveforms), with different THz polarizations.

## Appendix C: Hamiltonian for quadratic magnetorefractive effect

In order to describe the quadratic magnetorefractive effect observed in the present study, we consider the following Hamiltonian representing a nonlinear interaction between the spin dynamics and the (probe) electric field [2, 3]:

$$\mathcal{H}' = \frac{1}{2} \sum_{\xi,\xi'=x,y,z} E_\xi(t) E_{\xi'}(t) \left\{ \alpha_{\xi,\xi'} \boldsymbol{S}_1(t) \cdot \boldsymbol{S}_2(t) + \boldsymbol{\beta}_{\xi,\xi'} \cdot [\boldsymbol{S}_1(t) \times \boldsymbol{S}_2(t)] \right\}. \tag{C1}$$

Here, $E_\xi$ is the electric field in the $\xi = \{x, y, z\}$ direction. $\boldsymbol{S}_{1,2}$ are two $Fe^{3+}$ spins in the two-sublattice model. $\alpha_{\xi,\xi'}$ and $\boldsymbol{\beta}_{\xi,\xi'}$ are coefficients for isotropic and antisymmetric exchange interactions, respectively. Because the resonance of the charge-transfer transition from $O^{2-}$ to $Fe^{3+}$ places around the probe wavelength 800 nm in orthoferrites, it looks quite natural that the exchange interactions, which originates from the virtual charge transfers, are influenced by the electric field $E_\xi$, i.e., the isotropic exchange interaction strength $J_{Fe}$ and antisymmetric one $D_{Fe}$ are dynamically modulated by $E_\xi(t)$ [3].

On the other hand, in the present study, $\mathcal{H}'$ describes another phenomenon; the permittivity for the probe light is modified by the spin dynamics $\boldsymbol{S}_{1,2}(t)$. As discussed in the main text, we can rewrite Eq. (C1) as $\mathcal{H}' = \sum_{\xi,\xi'} \varepsilon_{\xi,\xi'}(t) E_\xi(t) E_{\xi'}(t)/2$ with the permittivity tensor

$$\varepsilon_{\xi,\xi'}(t) = \alpha_{\xi,\xi'} \boldsymbol{S}_1(t) \cdot \boldsymbol{S}_2(t) + \boldsymbol{\beta}_{\xi,\xi'} \cdot [\boldsymbol{S}_1(t) \times \boldsymbol{S}_2(t)]. \tag{C2}$$

The quadratic magnetorefractive effect observed in the present study is understood through this permittivity.

## Appendix D: Magnon excitation by THz pump

Since $YFeO_3$ and $ErFeO_3$ are in the $\Gamma_4$ phase in our experimental setup (at room temperature), the two $Fe^{3+}$ spins $\boldsymbol{S}_{1,2}$ are antiferromagnetically ordered along the $x$ (a) axis but slightly canted toward the $z$ (c) axis as depicted in Fig. 3. Obeying the two-sublattice model in Ref. [4], we consider the Hamiltonian of the $Fe^{3+}$ spins as

$$\mathcal{H}_{Fe} = zJ_{Fe}\boldsymbol{S}_1 \cdot \boldsymbol{S}_2 - zD_{Fe}(S_{1x}S_{2z} - S_{1z}S_{2x}) - A_x({S_{1x}}^2 + {S_{2x}}^2) - A_z({S_{1z}}^2 + {S_{2z}}^2) - A_{xz}(S_{1x}S_{1z} - S_{2x}S_{2z}). \tag{D1}$$

Here, $z = 6$ is the number of nearest neighboring $Fe^{3+}$ sites. $J_{Fe}$ and $D_{Fe}$ are isotropic and antisymmetric exchange interaction strengths. $A_{x,z,xz}$ are magnetic anisotropy energies. We use the following parameters for $YFeO_3$:

$$J_{Fe} = 4.96 \text{ meV}, \tag{D2a}$$
$$D_{Fe} = 0.107 \text{ meV}, \tag{D2b}$$
$$A_x = 0.00288 \text{ meV}, \tag{D2c}$$
$$A_z = 0.0008 \text{ meV}, \tag{D2d}$$
$$A_{xz} = 0. \tag{D2e}$$

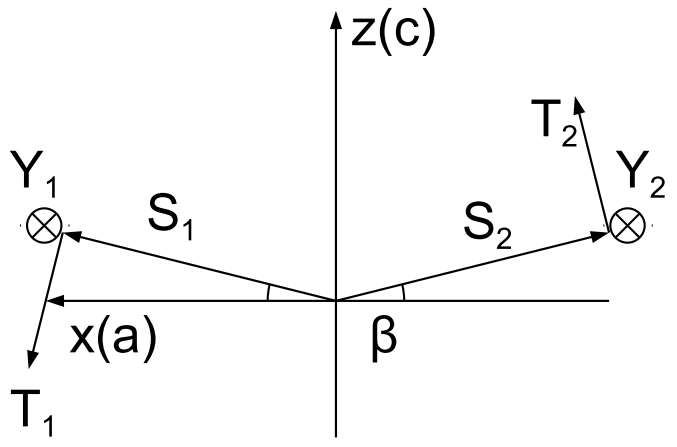

FIG. 3. Spins in the two-sublattice model of $YFeO_3$ and $ErFeO_3$.

The Hamiltonian for interaction between the spins $\boldsymbol{S}_{1,2}$ and external magnetic flux density $\boldsymbol{B}_{\rm ext}(t)$ (THz pump) is described as

$$\mathcal{H}_{\rm ext} = g\mu_{\rm B} \sum_{j=1,2} \boldsymbol{S}_j \cdot \boldsymbol{B}_{\rm ext}(t), \tag{D3}$$

where $g$ is the $g$-factor and $\mu_{\rm B}$ is the Bohr magneton. The equation of motion of $\boldsymbol{S}_{1,2}$ for Hamiltonian $\mathcal{H}_{\rm Fe} + \mathcal{H}_{\rm ext}$ is expressed as

$$\hbar\dot{\boldsymbol{S}}_j = -\boldsymbol{S}_j \times \left[g\mu_{\rm B}\boldsymbol{B}_{\rm ext}(t) + \boldsymbol{\nabla}_{\boldsymbol{S}_j}\mathcal{H}_{\rm Fe}\right], \tag{D4a}$$

$$\gamma^{-1}\dot{\boldsymbol{S}}_j = -\boldsymbol{S}_j \times \left[\boldsymbol{B}_{\rm ext}(t) + \boldsymbol{\nabla}_{\boldsymbol{S}_j}\mathcal{H}_{\rm Fe}/(g\mu_{\rm B})\right], \tag{D4b}$$

where $\gamma = g\mu_{\rm B}/\hbar$ is the gyromagnetic ratio. The gradients are derived as

$$\boldsymbol{\nabla}_{\boldsymbol{S}_1}\mathcal{H}_{\rm Fe} = \begin{pmatrix} zJ_{\rm Fe}S_{2x} - zD_{\rm Fe}S_{2z} - 2A_xS_{1x} - A_{xz}S_{1z} \\ zJ_{\rm Fe}S_{2y} \\ zJ_{\rm Fe}S_{2z} + zD_{\rm Fe}S_{2x} - 2A_zS_{1z} - A_{xz}S_{1x} \end{pmatrix}, \tag{D5a}$$

$$\boldsymbol{\nabla}_{\boldsymbol{S}_2}\mathcal{H}_{\rm Fe} = \begin{pmatrix} zJ_{\rm Fe}S_{1x} + zD_{\rm Fe}S_{1z} - 2A_xS_{2x} + A_{xz}S_{2z} \\ zJ_{\rm Fe}S_{1y} \\ zJ_{\rm Fe}S_{1z} - zD_{\rm Fe}S_{1x} - 2A_zS_{2z} + A_{xz}S_{2x} \end{pmatrix}. \tag{D5b}$$

The spins $\bar{\boldsymbol{S}}_{1,2}$ in the absence of the external magnetic field (THz pump) are determined by

$$\bar{\boldsymbol{S}}_j \times \overline{\boldsymbol{\nabla}_{\boldsymbol{S}_j}\mathcal{H}_{\rm Fe}} = \boldsymbol{0}, \tag{D6}$$

where the over bar means an equilibrium (static; non-oscillating) component. As depicted in Fig. 3, we represent the equilibrium spins by the canting angle $\beta_0$ and $Fe^{3+}$ spin amplitude $S = 5/2$ as

$$\bar{\boldsymbol{S}}_1 = S\begin{pmatrix}\cos\beta_0 \\ 0 \\ \sin\beta_0\end{pmatrix}, \quad \bar{\boldsymbol{S}}_2 = S\begin{pmatrix}-\cos\beta_0 \\ 0 \\ \sin\beta_0\end{pmatrix}. \tag{D7}$$

From Eq. (D6), the canting angle $\beta_0$ is determined by

$$\tan(2\beta_0) = \frac{zD_{\rm Fe} + A_{xz}}{zJ_{\rm Fe} + A_x - A_z}. \tag{D8}$$

Let us discuss the modulation dynamics of spins from the equilibrium, i.e., magnons. We define the spin modulations as

$$\delta\boldsymbol{S}_j(t) \equiv \boldsymbol{S}_j(t) - \bar{\boldsymbol{S}}_j \tag{D9}$$

Following Ref. [4], we approximate the spin modulations by newly introducing $\{T_{1,2}, Y_{1,2}\}$ as depicted in Fig. 3:

$$\delta\boldsymbol{S}_1(t) \approx \begin{bmatrix} T_1(t)\sin\beta_0 \\ Y_1(t) \\ -T_1(t)\cos\beta_0 \end{bmatrix}, \quad \delta\boldsymbol{S}_2(t) \approx \begin{bmatrix} T_2(t)\sin\beta_0 \\ Y_2(t) \\ T_2(t)\cos\beta_0 \end{bmatrix} \tag{D10}$$

This approximation is justified when the modulations $\delta \boldsymbol{S}_j(t)$ are much smaller than the equilibrium vectors $\bar{\boldsymbol{S}}_j$. From Eq. (D4), the equation of motion of the spin modulations are rewritten as

$$\gamma^{-1}\delta\dot{\boldsymbol{S}}_j = -\delta\left\{\boldsymbol{S}_j(t)\times\boldsymbol{\nabla}_{\boldsymbol{S}_j}\mathcal{H}_{\rm Fe}/(g\mu_{\rm B})\right\} - \bar{\boldsymbol{S}}_j\times\boldsymbol{B}_{\rm ext}(t). \tag{D11a}$$

$$\gamma^{-1}\frac{\partial}{\partial t}\begin{bmatrix} T_1(t)\sin\beta_0 \\ Y_1(t) \\ -T_1(t)\cos\beta_0\end{bmatrix} \approx -\delta\left\{\boldsymbol{S}_1(t)\times\boldsymbol{\nabla}_{\boldsymbol{S}_1}\mathcal{H}_{\rm Fe}/(g\mu_{\rm B})\right\} - S\begin{bmatrix} -B_{{\rm ext},y}(t)\sin\beta_0 \\ B_{{\rm ext},x}(t)\sin\beta_0 - B_{{\rm ext},z}(t)\cos\beta_0 \\ B_{{\rm ext},y}(t)\cos\beta_0\end{bmatrix}, \tag{D11b}$$

$$\gamma^{-1}\frac{\partial}{\partial t}\begin{bmatrix} T_2(t)\sin\beta_0 \\ Y_2(t) \\ T_2(t)\cos\beta_0\end{bmatrix} \approx -\delta\left\{\boldsymbol{S}_2(t)\times\boldsymbol{\nabla}_{\boldsymbol{S}_2}\mathcal{H}_{\rm Fe}/(g\mu_{\rm B})\right\} - S\begin{bmatrix} -B_{{\rm ext},y}(t)\sin\beta_0 \\ B_{{\rm ext},x}(t)\sin\beta_0 + B_{{\rm ext},z}(t)\cos\beta_0 \\ -B_{{\rm ext},y}(t)\cos\beta_0\end{bmatrix}. \tag{D11c}$$

The first terms were discussed in Refs. [4, 5], and they eventually give the first two terms in the following four equations for $T_{1,2}$ and $Y_{1,2}$:

$$\gamma^{-1}\dot{T}_1(t) = -aY_1(t) + bY_2(t) + SB_{{\rm ext},y}(t), \tag{D12a}$$

$$\gamma^{-1}\dot{Y}_1(t) = -cT_1(t) - dT_2(t) - S[B_{{\rm ext},x}(t)\sin\beta_0 - B_{{\rm ext},z}(t)\cos\beta_0], \tag{D12b}$$

$$\gamma^{-1}\dot{T}_2(t) = -aY_2(t) + bY_1(t) + SB_{{\rm ext},y}(t), \tag{D12c}$$

$$\gamma^{-1}\dot{Y}_2(t) = -cT_2(t) - dT_1(t) - S[B_{{\rm ext},x}(t)\sin\beta_0 + B_{{\rm ext},z}(t)\cos\beta_0]. \tag{D12d}$$

The last terms are derived by comparing the left-hand side and the last terms on the right-hand side in Eqs. (D11). As discussed in Refs. [4, 5], the coefficients $\{a, b, c, d\}$ are expressed as

$$a = [S/(g\mu_{\rm B})]\left[-A_x - A_z - (zJ_{\rm Fe} + A_x - A_z)\cos(2\beta_0) - (A_{xz} + zD_{\rm Fe})\sin(2\beta_0)\right], \tag{D13a}$$

$$b = [S/(g\mu_{\rm B})]zJ_{\rm Fe}, \tag{D13b}$$

$$c = [S/(g\mu_{\rm B})]\left[(zJ_{\rm Fe} + 2A_x - 2A_z)\cos(2\beta_0) - zD_{\rm Fe}\sin(2\beta_0)\right], \tag{D13c}$$

$$d = [S/(g\mu_{\rm B})]\left[-zJ_{\rm Fe}\cos(2\beta_0) + (2A_{xz} + zD_{\rm Fe})\sin(2\beta_0)\right], \tag{D13d}$$

$$\begin{aligned} b + a &= [S/(g\mu_{\rm B})]\left[zJ_{\rm Fe} - A_x - A_z - (zJ_{\rm Fe} + A_x - A_z)\cos(2\beta_0) - (A_{xz} + zD_{\rm Fe})\sin(2\beta_0)\right] \\ &\approx [S/(g\mu_{\rm B})]\left[-(z\bar{J}_{\rm Fe} - zJ_{\rm Fe}) - 2A_x \approx -(A_{xz} + zD_{\rm Fe})\tan\beta_0 - 2A_x\right], \end{aligned} \tag{D14a}$$

$$\begin{aligned} b - a &= [S/(g\mu_{\rm B})]\left[zJ_{\rm Fe} + A_x + A_z + (zJ_{\rm Fe} + A_x - A_z)\cos(2\beta_0) + (A_{xz} + zD_{\rm Fe})\sin(2\beta_0)\right] \\ &\approx [S/(g\mu_{\rm B})]2zJ_{\rm Fe}, \end{aligned} \tag{D14b}$$

$$\begin{aligned} d + c &= [S/(g\mu_{\rm B})]\left[2(A_x - A_z)\cos(2\beta_0) + 2A_{xz}\sin(2\beta_0)\right] \\ &\approx [S/(g\mu_{\rm B})]2(A_x - A_z), \end{aligned} \tag{D14c}$$

$$\begin{aligned} d - c &= [S/(g\mu_{\rm B})]\left[-2(zJ_{\rm Fe} + A_x - A_z)\cos(2\beta_0) + 2(A_{xz} + zD_{\rm Fe})\sin(2\beta_0)\right] \\ &\approx [S/(g\mu_{\rm B})]\left[-2zJ_{\rm Fe}\right], \end{aligned} \tag{D14d}$$

where we defined

$$\bar{J}_{\rm Fe} \equiv J_{\rm Fe}\cos(2\beta_0) + (A_{xz}/z + D_{\rm Fe})\sin(2\beta_0). \tag{D15}$$

Here, we newly define

$$T_\pm \equiv T_1 \pm T_2, \quad Y_\pm \equiv Y_1 \pm Y_2. \tag{D16}$$

In terms of them, the spin modulations are expressed as

$$\delta\boldsymbol{S}_1(t) \approx \frac{1}{2}\begin{bmatrix}(T_+ + T_-)\sin\beta_0 \\ (Y_+ + Y_-) \\ -(T_+ + T_-)\cos\beta_0\end{bmatrix}, \quad \delta\boldsymbol{S}_2(t) \approx \frac{1}{2}\begin{bmatrix}(T_+ - T_-)\sin\beta_0 \\ (Y_+ - Y_-) \\ (T_+ - T_-)\cos\beta_0\end{bmatrix}. \tag{D17}$$

The "+" and "−" modes correspond to the quasi-ferromagnetic and quasi-antiferromagnetic modes (FM and AFM), respectively. From Eqs. (D12), the equations of motion of them are derived as

$$\gamma^{-1}\dot{T}_+(t) = (b - a)Y_+(t) + 2SB_{{\rm ext},y}(t), \tag{D18a}$$

$$\gamma^{-1}\dot{Y}_+(t) = -(d + c)T_+(t) - 2SB_{{\rm ext},x}(t)\sin\beta_0, \tag{D18b}$$

$$\gamma^{-1}\dot{T}_-(t) = -(b + a)Y_-(t), \tag{D18c}$$

$$\gamma^{-1}\dot{Y}_-(t) = (d - c)T_-(t) + 2SB_{{\rm ext},z}(t)\cos\beta_0. \tag{D18d}$$

For the $\omega$-Fourier transform $\tilde{T}_\pm, \tilde{Y}_\pm$ of them, such as

$$T_\pm(t) = \mathrm{Re}[\tilde{T}_\pm \mathrm{e}^{-\mathrm{i}\omega t}], \tag{D19}$$

these equations are solved as

$$\begin{pmatrix} \tilde{T}_+ \\ \tilde{Y}_+ \end{pmatrix} = \frac{2S\gamma}{\omega^2 - {\omega_+}^2} \begin{bmatrix} \mathrm{i}\omega & -\gamma(b-a) \\ \gamma(d+c) & \mathrm{i}\omega \end{bmatrix} \begin{pmatrix} \tilde{B}_{\mathrm{ext},y} \\ -\tilde{B}_{\mathrm{ext},x}\sin\beta_0 \end{pmatrix}, \tag{D20a}$$

$$\begin{pmatrix} \tilde{T}_- \\ \tilde{Y}_- \end{pmatrix} = \frac{2S\gamma\cos\beta_0}{\omega^2 - {\omega_-}^2} \begin{bmatrix} \gamma(b+a) \\ \mathrm{i}\omega \end{bmatrix} \tilde{B}_{\mathrm{ext},z}. \tag{D20b}$$

Here, $\omega_\pm$ are the eigenfrequencies given by

$${\omega_\pm}^2 = \gamma^2(b \mp a)(d \pm c). \tag{D21}$$

In the present study, the pump THz wave propagates along the $z$ axis. As seen in the above expression, the phase of the FM ("+") magnon oscillation is shifted by $\pi$ between $\tilde{B}_{\mathrm{ext},y}$ ($E \parallel x$) and $\tilde{B}_{\mathrm{ext},x}$ ($E \parallel y$). On the other hand, since $\tilde{B}_{\mathrm{ext},z} = 0$, the AFM ("−") magnons are basically not excited in our setup.

### Appendix E: Magnon oscillation in permittivity

Let us discuss the permittivity modulation in Eq. (C2) from the spin dynamics discussed in Appendix D. Here, following our setup, we assume $\tilde{B}_{\mathrm{ext},x} = \tilde{B}_{\mathrm{ext},z} = 0$. From Eqs. (D20), the FM's $\omega$-Fourier components of $T_+$ and $Y_+$ are obtained as

$$\tilde{T}_+ = \frac{\mathrm{i}\omega\gamma}{\omega^2 - {\omega_+}^2} 2S\tilde{B}_{\mathrm{ext}}, \quad \tilde{Y}_+ = \frac{\gamma(d+c)}{\mathrm{i}\omega}\tilde{T}_+ \tag{E1}$$

From Eqs. (D17), the $\omega$-Fourier components of spin modulations are expressed as

$$\delta\tilde{\boldsymbol{S}}_1 \approx \frac{\tilde{T}_+}{2} \begin{pmatrix} \sin\beta_0 \\ \frac{\gamma(d+c)}{\mathrm{i}\omega} \\ -\cos\beta_0 \end{pmatrix}, \quad \delta\tilde{\boldsymbol{S}}_2 \approx \frac{\tilde{T}_+}{2} \begin{pmatrix} \sin\beta_0 \\ \frac{\gamma(d+c)}{\mathrm{i}\omega} \\ \cos\beta_0 \end{pmatrix}. \tag{E2}$$

In Eq. (C2), the $\omega$- and $2\omega$-Fourier components of the permittivity modulation are expressed, respectively, as

$$\tilde{\varepsilon}_{\xi,\xi'}(\omega) = \alpha_{\xi,\xi'}\left(\bar{\boldsymbol{S}}_1 \cdot \delta\tilde{\boldsymbol{S}}_2 + \delta\tilde{\boldsymbol{S}}_1 \cdot \bar{\boldsymbol{S}}_2\right) + \boldsymbol{\beta}_{\xi,\xi'} \cdot \left(\bar{\boldsymbol{S}}_1 \times \delta\tilde{\boldsymbol{S}}_2 + \delta\tilde{\boldsymbol{S}}_1 \times \bar{\boldsymbol{S}}_2\right), \tag{E3a}$$

$$\tilde{\varepsilon}_{\xi,\xi'}(2\omega) = \alpha_{\xi,\xi'}\delta\tilde{\boldsymbol{S}}_1 \cdot \delta\tilde{\boldsymbol{S}}_2 + \boldsymbol{\beta}_{\xi,\xi'} \cdot \left(\delta\tilde{\boldsymbol{S}}_1 \times \delta\tilde{\boldsymbol{S}}_2\right). \tag{E3b}$$

From Eqs. (D7) and (E2), the products for the $\omega$-component are rewritten as

$$\bar{\boldsymbol{S}}_1 \cdot \delta\tilde{\boldsymbol{S}}_2 + \delta\tilde{\boldsymbol{S}}_1 \cdot \bar{\boldsymbol{S}}_2 = 0, \tag{E4a}$$

$$\bar{\boldsymbol{S}}_1 \times \delta\tilde{\boldsymbol{S}}_2 + \delta\tilde{\boldsymbol{S}}_1 \times \bar{\boldsymbol{S}}_2 = S\cos\beta_0 \frac{\gamma(d+c)}{\mathrm{i}\omega} \begin{pmatrix} 0 \\ 0 \\ 1 \end{pmatrix} \tilde{T}_+. \tag{E4b}$$

In this way, the $\omega$-component depends only on $\beta_{\xi,\xi',z}$. This is because the equilibrium ferromagnetic vector $\bar{\boldsymbol{S}}_1 + \bar{\boldsymbol{S}}_2$ and the dynamical one $\delta\tilde{\boldsymbol{S}}_1 + \delta\tilde{\boldsymbol{S}}_2(t)$ are perpendicular to each other during the FM precession and $\bar{\boldsymbol{S}}_j \cdot \tilde{\boldsymbol{S}}_j = \boldsymbol{0}$. On the other hand, the products for the $2\omega$-component are

$$\delta\tilde{\boldsymbol{S}}_1 \cdot \delta\tilde{\boldsymbol{S}}_2 = -\frac{{\tilde{T}_+}^2}{4}\left[\cos(2\beta_0) + \frac{\gamma^2(d+c)^2}{\omega^2}\right], \tag{E5a}$$

$$\delta\tilde{\boldsymbol{S}}_1 \times \delta\tilde{\boldsymbol{S}}_2 = \frac{{\tilde{T}_+}^2}{4} \begin{bmatrix} 2\frac{\gamma(d+c)}{\mathrm{i}\omega}\cos\beta_0 \\ -\sin(2\beta_0) \\ 0 \end{bmatrix}. \tag{E5b}$$

In this way, the $2\omega$-component depend on $\alpha_{\xi,\xi'}$, $\beta_{\xi,\xi',x}$, and $\beta_{\xi,\xi',y}$. If we can assume $\alpha_{\xi,\xi'} \gg \beta_{\xi,\xi',x,y}$, which is usually justified for standard isotropic and antisymmetric exchange interactions, we can consider that the $2\omega$-component mostly reflects $\alpha_{\xi,\xi'}$. Since the electric field of the probe light is along the $x$ (a) axis in our setup, our experimental results basically reflects $\varepsilon_{x,x}$, $\alpha_{x,x}$, and $\boldsymbol{\beta}_{x,x}$.

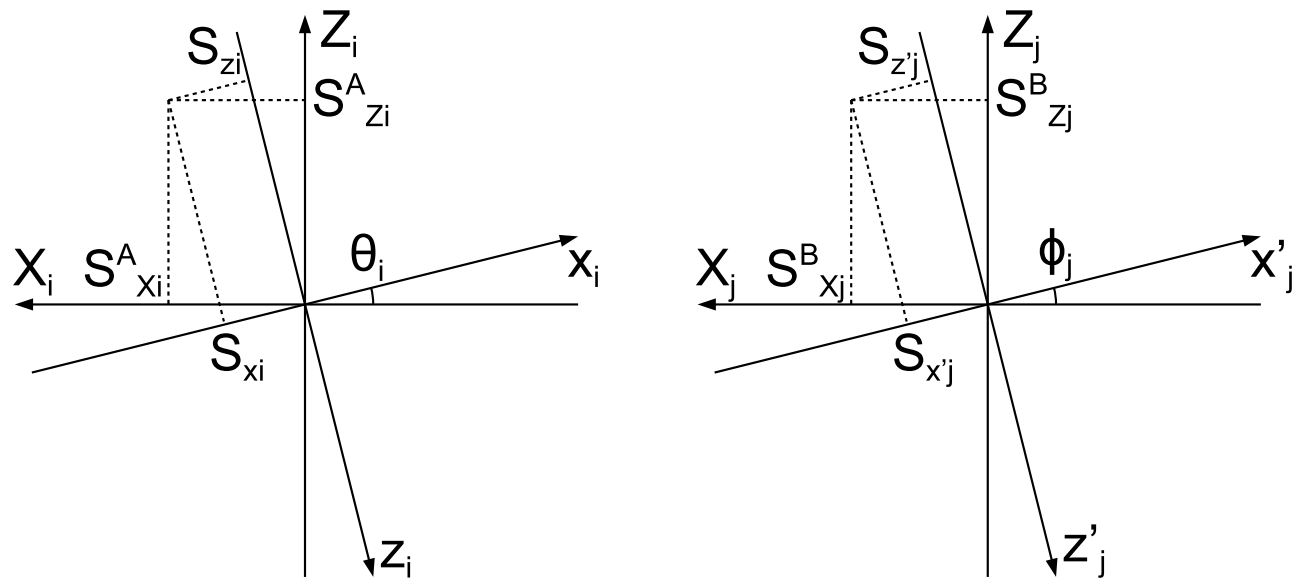

FIG. 4. Axes for calculating the FM and AFM magnon modes

## Appendix F: Nonlinear coupling between AFM and FM magnons

Here, we will see analytically that a AFM magnon can be excited by two FM magnons by extending the calculation by Tsang, White, and White in Appendix of Ref. [5]. We consider the following Hamiltonian:

$$\mathcal{H}_0 = 2J_{\rm Fe}\sum_{\rm n.n.}\boldsymbol{S}_i^A\cdot\boldsymbol{S}_j^B + D_{\rm Fe}\sum_{\rm n.n.}(S^A_{Z_i}S^B_{X_j} - S^A_{X_i}S^B_{Z_j}) - A_z\left(\sum_i {S^A_{Z_i}}^2 + \sum_j {S^B_{Z_j}}^2\right) - A_x\left(\sum_i {S^A_{X_i}}^2 + \sum_j {S^B_{X_j}}^2\right). \tag{F1}$$

Here, $\boldsymbol{S}_i^A$ and $\boldsymbol{S}_j^B$ are dimensionless spins at A and B sub-lattices of $Fe^{3+}$, respectively. $\sum_{\rm n.n.}$ means a summation over all the nearest neighbor couplings. We additionally consider an external magnetic flux density $\boldsymbol{B}_{\rm ext}$. The total Hamiltonian is

$$\mathcal{H} = \mathcal{H} + \mathcal{H}_{\rm ext}. \tag{F2}$$

Here, the interaction Hamiltonian $\mathcal{H}_{\rm ext}$ between the spins and external magnetic field is

$$\mathcal{H}_{\rm ext} \equiv -\boldsymbol{\mu}\cdot\boldsymbol{B}_{\rm ext} = g\mu_{\rm B}\left(\sum_i \boldsymbol{S}_i^A + \sum_j \boldsymbol{S}_j^B\right)\cdot\boldsymbol{B}_{\rm ext} = \left(\sum_i \boldsymbol{S}_i^A + \sum_j \boldsymbol{S}_j^B\right)\cdot\boldsymbol{\mathcal{B}}_{\rm ext}, \tag{F3}$$

where we defined

$$\boldsymbol{\mathcal{B}}_{\rm ext} \equiv g\mu_{\rm B}\boldsymbol{B}_{\rm ext}. \tag{F4}$$

This has the dimension of energy. $g$ is the (Lande's) g-factor, $\mu_{\rm B} = e\hbar/(2mc)$ is the Bohr magneton.

As shown in Fig. 4, we rewrite $\boldsymbol{S}_i^A$ and $\boldsymbol{S}_j^B$ from the original $\{X, Y, Z\}$ basis into the $\{x_i, y_i = Y, z_i\}$ basis rotated by $\pi - \theta_i$ and into $\{x'_j, y'_j = Y, z'_j\}$ rotated by $\pi - \phi_j$, respectively.

$$S^A_{Z_i} = -S_{z_i}\cos\theta_i + S_{x_i}\sin\theta_i, \tag{F5a}$$

$$S^A_{X_i} = -S_{x_i}\cos\theta_i - S_{z_i}\sin\theta_i, \tag{F5b}$$

$$S^B_{Z_j} = -S_{z'_j}\cos\phi_j + S_{x'_j}\sin\phi_j, \tag{F5c}$$

$$S^B_{X_j} = -S_{x'_j}\cos\phi_j - S_{z'_j}\sin\phi_j. \tag{F5d}$$

The terms in Eq. (F1) are rewritten as

$$S^A_{Z_i}S^B_{Z_j} + S^A_{X_i}S^B_{X_j} = (S_{z_i}S_{z'_j} + S_{x_i}S_{x'_j})\cos(\theta_i - \phi_j) + (S_{z_i}S_{x'_j} - S_{x_i}S_{z'_j})\sin(\theta_i - \phi_j), \tag{F6a}$$

$$S^A_{Z_i}S^B_{X_j} - S^A_{X_i}S^B_{Z_j} = -(S_{z_i}S_{z'_j} + S_{x_i}S_{x'_j})\sin(\theta_i - \phi_j) + (S_{z_i}S_{x'_j} - S_{x_i}S_{z'_j})\cos(\theta_i - \phi_j), \tag{F6b}$$

$$A_z{S^A_{Z_i}}^2 + A_x{S^A_{X_i}}^2 = {S_{z_i}}^2(A_z\cos^2\theta_i + A_x\sin^2\theta_i) + {S_{x_i}}^2(A_z\sin^2\theta_i + A_x\cos^2\theta_i) + S_{z_i}S_{x_i}(A_x - A_z)\sin(2\theta_i), \tag{F7a}$$

$$A_z{S^B_{Z_j}}^2 + A_x{S^B_{X_j}}^2 = {S_{z'_j}}^2(A_z\cos^2\phi_j + A_x\sin^2\phi_j) + {S_{x'_j}}^2(A_z\sin^2\phi_j + A_x\cos^2\phi_j) + S_{z'_j}S_{x'_j}(A_x - A_z)\sin(2\phi_j). \tag{F7b}$$

Then, the Hamiltonian in Eq. (F1) is rewritten as

$$\begin{aligned}\mathcal{H}_0 = & 2J_{\rm Fe}\sum_{\rm n.n.}\Big[(S_{z_i}S_{z'_j}+S_{x_i}S_{x'_j})\cos(\theta_i-\phi_j)+(S_{z_i}S_{x'_j}-S_{x_i}S_{z'_j})\sin(\theta_i-\phi_j)+S_{y_i}S_{y'_j}\Big]\\ &+D_{\rm Fe}\sum_{\rm n.n.}\Big[-(S_{z_i}S_{z'_j}+S_{x_i}S_{x'_j})\sin(\theta_i-\phi_j)+(S_{z_i}S_{x'_j}-S_{x_i}S_{z'_j})\cos(\theta_i-\phi_j)\Big]\\ &-\sum_i\left[{S_{z_i}}^2(A_z\cos^2\theta_i+A_x\sin^2\theta_i)+{S_{x_i}}^2(A_z\sin^2\theta_i+A_x\cos^2\theta_i)+S_{z_i}S_{x_i}(A_x-A_z)\sin(2\theta_i)\right]\\ &-\sum_j\Big[{S_{z'_j}}^2(A_z\cos^2\phi_j+A_x\sin^2\phi_j)+{S_{x'_j}}^2(A_z\sin^2\phi_j+A_x\cos^2\phi_j)+S_{z'_j}S_{x'_j}(A_x-A_z)\sin(2\phi_j)\Big]. \end{aligned} \tag{F8}$$

Although the last terms proportional to $(A_x - A_z)$ are different from those in the paper of Tsang, White, and White [5], those terms will not appear in the following calculations.

Here, we assume that the external magnetic flux density $\boldsymbol{B}_{\rm ext}$ is in the $x-z$ plane, and it is characterized with an angle $\alpha$ as

$$\mathcal{B}^z_{\rm ext} \equiv \mathcal{B}_{\rm ext}\sin\alpha, \tag{F9a}$$
$$\mathcal{B}^x_{\rm ext} \equiv -\mathcal{B}_{\rm ext}\cos\alpha. \tag{F9b}$$

Then, $\mathcal{H}_{\rm ext}$ is rewritten as

$$\mathcal{H}_{\rm ext} = \left(\sum_i S^A_{Z_i}+\sum_j S^B_{Z_j}\right)\mathcal{B}^z_{\rm ext}+\left(\sum_i S^A_{X_i}+\sum_j S^B_{X_j}\right)\mathcal{B}^x_{\rm ext} \tag{F10a}$$
$$= \sum_i\left[S_{z_i}\sin(\theta_i-\alpha)+S_{x_i}\cos(\theta_i-\alpha)\right]\mathcal{B}_{\rm ext}+\sum_j\Big[S_{z'_j}\sin(\phi_j-\alpha)+S_{x'_j}\cos(\phi_j-\alpha)\Big]\mathcal{B}_{\rm ext}. \tag{F10b}$$

The stable state (ground state) is obtained as follows. Assuming $S_{x_{i,j}} = S$, $S_{y_{i,j}} = 0$, and $S_{z_{i,j}} = 0$, i.e., without the fluctuations, the Hamiltonian in Eq. (F8) and (F10) are rewritten as

$$\mathcal{H}_0/S^2 = \sum_{\rm n.n.}\left[2J_{\rm Fe}\cos(\theta_i-\phi_j)-D_{\rm Fe}\sin(\theta_i-\phi_j)\right]-\sum_i(A_z\sin^2\theta_i+A_x\cos^2\theta_i)-\sum_j(A_z\sin^2\phi_j+A_x\cos^2\phi_j), \tag{F11}$$

$$\mathcal{H}_{\rm ext}/S = \sum_i\cos(\theta_i-\alpha)\mathcal{B}_{\rm ext}+\sum_j\cos(\phi_j-\alpha)\mathcal{B}_{\rm ext}. \tag{F12}$$

Assuming that all the spins in each sub-lattice are ordered along the same direction as

$$\theta_i = \theta, \tag{F13a}$$
$$\phi_j = \phi, \tag{F13b}$$

the Hamiltonian is rewritten as

$$\begin{aligned}\mathcal{H}/(NS^2) = & \left[2zJ_{\rm Fe}\cos(\theta-\phi)-zD_{\rm Fe}\sin(\theta-\phi)\right]-A_z(\sin^2\theta+\sin^2\phi)-A_x(\cos^2\theta+\cos^2\phi)\\ &+\left[\cos(\theta-\alpha)+\cos(\phi-\alpha)\right](\mathcal{B}_{\rm ext}/S). \end{aligned} \tag{F14}$$

Here, the nearest neighbor summation is replaced by $zN$, where $N$ is the number of all the $Fe^{3+}$ ions in each sub-lattice and $z = 6$ is the number of nearest neighbor $Fe^{3+}$ spins. The stable angles $\theta$ and $\phi$ are obtained by solving

$$\frac{\partial\mathcal{H}}{\partial\theta} = -2zJ_{\rm Fe}\sin(\theta-\phi)-zD_{\rm Fe}\cos(\theta-\phi)-(A_z-A_x)\sin(2\theta)-\frac{\mathcal{B}_{\rm ext}}{S}\sin(\theta-\alpha)=0, \tag{F15a}$$
$$\frac{\partial\mathcal{H}}{\partial\phi} = 2zJ_{\rm Fe}\sin(\theta-\phi)+zD_{\rm Fe}\cos(\theta-\phi)-(A_z-A_x)\sin(2\phi)-\frac{\mathcal{B}_{\rm ext}}{S}\sin(\phi-\alpha)=0. \tag{F15b}$$

From the stable state where the spins are directed along $x_i$ and $x'_j$ directions as we considered above, the fluctuation in the $\{y_i, z_i\}$ and $\{y'_j, z'_j\}$ directions obey the following equations obtained from Eqs. (F8) and (F10):

$$\begin{aligned}
\hbar\dot{S}_{x_i} =& \sum_{j=\mathrm{n.n.}i} \Big\{-S_{y_i}S_{z'_j}\left[2J_{\mathrm{Fe}}\cos(\theta_i-\phi_j)-D_{\mathrm{Fe}}\sin(\theta_i-\phi_j)\right]-S_{y_i}S_{x'_j}\left[2J_{\mathrm{Fe}}\sin(\theta_i-\phi_j)+D_{\mathrm{Fe}}\cos(\theta_i-\phi_j)\right]+2J_{\mathrm{Fe}}S_{z_i}S_{y'_j}\Big\} \\
&+(S_{y_i}S_{z_i}+S_{z_i}S_{y_i})(A_z\cos^2\theta_i+A_x\sin^2\theta_i)+S_{y_i}S_{x_i}(A_x-A_z)\sin(2\theta_i)-S_{y_i}\mathcal{B}_{\mathrm{ext}}\sin(\theta_i-\alpha), &\text{(F16a)}\\
\hbar\dot{S}_{x'_j} =& \sum_{i=\mathrm{n.n.}j} \Big\{-S_{z_i}S_{y'_j}\left[2J_{\mathrm{Fe}}\cos(\theta_i-\phi_j)-D_{\mathrm{Fe}}\sin(\theta_i-\phi_j)\right]+S_{x_i}S_{y'_j}\left[2J_{\mathrm{Fe}}\sin(\theta_i-\phi_j)+D_{\mathrm{Fe}}\cos(\theta_i-\phi_j)\right]+2J_{\mathrm{Fe}}S_{y_i}S_{z'_j}\Big\} \\
&+(S_{y'_j}S_{z'_j}+S_{z'_j}S_{y'_j})(A_z\cos^2\phi_j+A_x\sin^2\phi_j)+S_{y'_j}S_{x'_j}(K_y-A_x)\sin(2\phi_j)-S_{y'_j}\mathcal{B}_{\mathrm{ext}}\sin(\phi_j-\alpha), &\text{(F16b)}\\
\hbar\dot{S}_{y_i} =& \sum_{j=\mathrm{n.n.}i}(S_{x_i}S_{z'_j}-S_{z_i}S_{x'_j})\left[2J_{\mathrm{Fe}}\cos(\theta_i-\phi_j)-D_{\mathrm{Fe}}\sin(\theta_i-\phi_j)\right] \\
&+\sum_{j=\mathrm{n.n.}i}(S_{x_i}S_{x'_j}+S_{z_i}S_{z'_j})\left[2J_{\mathrm{Fe}}\sin(\theta_i-\phi_j)+D_{\mathrm{Fe}}\cos(\theta_i-\phi_j)\right] \\
&-(S_{x_i}S_{z_i}+S_{z_i}S_{x_i})(A_z\cos^2\theta_i+A_x\sin^2\theta_i)+(S_{z_i}S_{x_i}+S_{x_i}S_{z_i})(A_z\sin^2\theta_i+A_x\cos^2\theta_i) \\
&-({S_{x_i}}^2-{S_{z_i}}^2)(A_x-A_z)\sin(2\theta_i)+S_{x_i}\mathcal{B}_{\mathrm{ext}}\sin(\theta_i-\alpha)-S_{z_i}\mathcal{B}_{\mathrm{ext}}\cos(\theta_i-\alpha), &\text{(F16c)}\\
\hbar\dot{S}_{y'_j} =& \sum_{i=\mathrm{n.n.}j}(S_{x'_j}S_{z_i}-S_{z'_j}S_{x_i})\left[2J_{\mathrm{Fe}}\cos(\theta_i-\phi_j)-D_{\mathrm{Fe}}\sin(\theta_i-\phi_j)\right] \\
&-\sum_{i=\mathrm{n.n.}j}(S_{x'_j}S_{x_i}+S_{z'_j}S_{z_i})\left[2J_{\mathrm{Fe}}\sin(\theta_i-\phi_j)+D_{\mathrm{Fe}}\cos(\theta_i-\phi_j)\right] \\
&-(S_{x'_j}S_{z'_j}+S_{z'_j}S_{x'_j})(A_z\cos^2\phi_j+A_x\sin^2\phi_j)+(S_{z'_j}S_{x'_j}+S_{x'_j}S_{z'_j})(A_z\sin^2\phi_j+A_x\cos^2\phi_j) \\
&-({S_{x'_j}}^2-{S_{z'_j}}^2)(A_x-A_z)\sin(2\phi_j)+S_{x'_j}\mathcal{B}_{\mathrm{ext}}\sin(\phi_j-\alpha)-S_{z'_j}\mathcal{B}_{\mathrm{ext}}\cos(\phi_j-\alpha), &\text{(F16d)}\\
\hbar\dot{S}_{z_i} =& \sum_{j=\mathrm{n.n.}i} \Big\{S_{y_i}S_{x'_j}\left[2J_{\mathrm{Fe}}\cos(\theta_i-\phi_j)-D_{\mathrm{Fe}}\sin(\theta_i-\phi_j)\right]-S_{y_i}S_{z'_j}\left[2J_{\mathrm{Fe}}\sin(\theta_i-\phi_j)+D_{\mathrm{Fe}}\cos(\theta_i-\phi_j)\right]-2J_{\mathrm{Fe}}S_{x_i}S_{y'_j}\Big\} \\
&-(S_{y_i}S_{x_i}+S_{x_i}S_{y_i})(A_z\sin^2\theta_i+A_x\cos^2\theta_i)-S_{z_i}S_{y_i}(A_x-A_z)\sin(2\theta_i)+S_{y_i}\mathcal{B}_{\mathrm{ext}}\cos(\theta_i-\alpha), &\text{(F16e)}\\
\hbar\dot{S}_{z'_j} =& \sum_{i=\mathrm{n.n.}j} \Big\{S_{x_i}S_{y'_j}\left[2J_{\mathrm{Fe}}\cos(\theta_i-\phi_j)-D_{\mathrm{Fe}}\sin(\theta_i-\phi_j)\right]+S_{z_i}S_{y'_j}\left[2J_{\mathrm{Fe}}\sin(\theta_i-\phi_j)+D_{\mathrm{Fe}}\cos(\theta_i-\phi_j)\right]-2J_{\mathrm{Fe}}S_{x'_j}S_{y_i}\Big\} \\
&-(S_{y'_j}S_{x'_j}+S_{x'_j}S_{y'_j})(A_z\sin^2\phi_j+A_x\cos^2\phi_j)-S_{z'_j}S_{y'_j}(A_x-A_z)\sin(2\phi_j)+S_{y'_j}\mathcal{B}_{\mathrm{ext}}\cos(\phi_j-\alpha). &\text{(F16f)}
\end{aligned}$$

### 1. Linear dynamics of magnons under static magnetic field

First, let us assume $S_{x_{i,j}} \sim S \gg S_{z_{i,j}}, S_{y_{i,j}}$. We also assume that all the spins in each sub-lattice are oriented in the same directions as $\theta_i \to \theta$ and $\phi_j \to \phi$, while general situations (with domain walls, etc.) are considered in the discussion of Tsang, White, and White [5]. Using Eqs. (F15), the above equations are approximated and reduced to

$$\begin{aligned}
\hbar\dot{S}_z/S &\approx -2zJ_{\mathrm{Fe}}S_{y'}-\left[2z\bar{J}_{\mathrm{Fe}}+2A_z+2(A_x-A_z)\cos^2\theta-(\mathcal{B}_{\mathrm{ext}}/S)\cos(\theta-\alpha)\right]S_y, &\text{(F17a)}\\
\hbar\dot{S}_{z'}/S &\approx -2zJ_{\mathrm{Fe}}S_{y}-\left[2z\bar{J}_{\mathrm{Fe}}+2A_z+2(A_x-A_z)\cos^2\phi-(\mathcal{B}_{\mathrm{ext}}/S)\cos(\phi-\alpha)\right]S_{y'}, &\text{(F17b)}\\
\hbar\dot{S}_y/S &\approx -2z\bar{J}_{\mathrm{Fe}}S_{z'}+\left[2z\bar{J}_{\mathrm{Fe}}+2(A_x-A_z)\cos(2\theta)-(\mathcal{B}_{\mathrm{ext}}/S)\cos(\theta-\alpha)\right]S_z, &\text{(F17c)}\\
\hbar\dot{S}_{y'}/S &\approx -2z\bar{J}_{\mathrm{Fe}}S_{z}+\left[2z\bar{J}_{\mathrm{Fe}}+2(A_x-A_z)\cos(2\phi)-(\mathcal{B}_{\mathrm{ext}}/S)\cos(\phi-\alpha)\right]S_{z'}, &\text{(F17d)}
\end{aligned}$$

where we defined

$$2\bar{J}_{\mathrm{Fe}} \equiv -2J_{\mathrm{Fe}}\cos(\theta-\phi)+D_{\mathrm{Fe}}\sin(\theta-\phi). \tag{F18}$$

The eigenfrequencies are obtained by solving

$$\hbar\frac{\partial}{\partial t}\begin{pmatrix}S_z\\S'_z\\S_y\\S'_y\end{pmatrix}=\begin{pmatrix}0&0&M_{1,3}&-2zJ_{\mathrm{Fe}}\\0&0&-2zJ_{\mathrm{Fe}}&M_{2,4}\\M_{3,1}&-2z\bar{J}_{\mathrm{Fe}}&0&0\\-2z\bar{J}_{\mathrm{Fe}}&M_{4,2}&0&0\end{pmatrix}\begin{pmatrix}S_z\\S'_z\\S_y\\S'_y\end{pmatrix}, \tag{F19}$$

where

$$M_{1,3} = -\left[2z\bar{J}_{\rm Fe} + 2A_z + 2(A_x - A_z)\cos^2\theta - (\mathcal{B}_{\rm ext}/S)\cos(\theta-\alpha)\right], \tag{F20a}$$

$$M_{2,4} = -\left[2z\bar{J}_{\rm Fe} + 2A_z + 2(A_x - A_z)\cos^2\phi - (\mathcal{B}_{\rm ext}/S)\cos(\phi-\alpha)\right], \tag{F20b}$$

$$M_{3,1} = 2z\bar{J}_{\rm Fe} + 2(A_x - A_z)\cos(2\theta) - (\mathcal{B}_{\rm ext}/S)\cos(\theta-\alpha), \tag{F20c}$$

$$M_{4,2} = 2z\bar{J}_{\rm Fe} + 2(A_x - A_z)\cos(2\phi) - (\mathcal{B}_{\rm ext}/S)\cos(\phi-\alpha). \tag{F20d}$$

### 2. Linear dynamics of magnons in absence of external magnetic field

Next, let us simply consider $\mathcal{B}_{\rm ext} = 0$. Here, we introduce the equilibrium canting angle $\beta_0$ as

$$\theta_i = \pi - \beta_0, \tag{F21a}$$

$$\phi_j = \beta_0. \tag{F21b}$$

Then, we get

$$\sin(\theta_i - \phi_j) = \sin(2\beta_0), \tag{F22a}$$

$$\cos(\theta_i - \phi_j) = -\cos(2\beta_0), \tag{F22b}$$

$$\sin(2\theta_i) = -\sin(2\beta_0), \tag{F22c}$$

$$\sin(2\phi_j) = \sin(2\beta_0), \tag{F22d}$$

(F22e)

$$2\bar{J}_{\rm Fe} = 2J_{\rm Fe}\cos(2\beta_0) + D_{\rm Fe}\sin(2\beta_0). \tag{F23}$$

From Eqs. (F15), $\beta_0$ is determined by

$$(2zJ_{\rm Fe} + A_x - A_z)\sin(2\beta_0) - zD_{\rm Fe}\cos(2\beta_0) = 0. \tag{F24}$$

Using this relation, instead of Eqs. (F17), Eqs. (F16) are approximated and reduced to

$$\hbar\dot{S}_z/S \approx -2zJ_{\rm Fe}S_{y'} - \left[2z\bar{J}_{\rm Fe} + 2A_z + 2(A_x - A_z)\cos^2\beta_0\right]S_y, \tag{F25a}$$

$$\hbar\dot{S}_{z'}/S \approx -2zJ_{\rm Fe}S_y - \left[2z\bar{J}_{\rm Fe} + 2A_z + 2(A_x - A_z)\cos^2\beta_0\right]S_{y'}, \tag{F25b}$$

$$\hbar\dot{S}_y/S \approx -2z\bar{J}_{\rm Fe}S_{z'} + \left[2z\bar{J}_{\rm Fe} + 2(A_x - A_z)\cos(2\beta_0)\right]S_z, \tag{F25c}$$

$$\hbar\dot{S}_{y'}/S \approx -2z\bar{J}_{\rm Fe}S_z + \left[2z\bar{J}_{\rm Fe} + 2(A_x - A_z)\cos(2\beta_0)\right]S_{z'}. \tag{F25d}$$

The eigenmodes can be categorized to the following two motions

$$\sigma_x \equiv S_x - S_{x'}, \qquad \sigma_y \equiv S_y + S_{y'}, \qquad \sigma_z \equiv S_z + S_{z'}, \tag{F26a}$$

$$\gamma_x \equiv S_x + S_{x'} - 2S, \qquad \gamma_y \equiv S_y - S_{y'}, \qquad \gamma_z \equiv S_z - S_{z'}. \tag{F26b}$$

The equations of these modes are obtained as

$$\hbar\dot{\sigma}_z/S = -\left[2z(\bar{J}_{\rm Fe} + J_{\rm Fe}) + 2A_z + 2(A_x - A_z)\cos^2\beta_0\right]\sigma_y, \tag{F27a}$$

$$\hbar\dot{\sigma}_y/S = 2(A_x - A_z)\cos(2\beta_0)\sigma_z, \tag{F27b}$$

$$\hbar\dot{\gamma}_z/S = -\left[2z(\bar{J}_{\rm Fe} - J_{\rm Fe}) + 2A_z + 2(A_x - A_z)\cos^2\beta_0\right]\gamma_y, \tag{F27c}$$

$$\hbar\dot{\gamma}_y/S = \left[4z\bar{J}_{\rm Fe} + 2(A_x - A_z)\cos(2\beta_0)\right]\gamma_z. \tag{F27d}$$

Here, assuming $J_{\rm Fe} \gg A_z, A_x$ and $\beta_0 \ll 1$, the eigenfrequencies of these modes are approximately expressed as

$$E_\sigma{}^2 \approx 4zJ_{\rm Fe}S[2(A_x - A_z)S], \tag{F28a}$$

$$E_\gamma{}^2 \approx 4zJ_{\rm Fe}S[2z(\bar{J}_{\rm Fe} - J_{\rm Fe})S + 2A_xS] = 4zJ_{\rm Fe}S[zD_{\rm Fe}S\tan\beta_0 + 2A_xS]. \tag{F28b}$$

They are the energies of free solutions (spin-wave-like solutions), Eqs. (A32) and (A34), in Appendix of Tsang, White, and White (1978) [5]. By replacing $\{x, y, z\} \to \{z, x, y\}$ and setting $\eta(k) = 1$, they are also equivalent with Eqs. (24)

of Tsang, White, and Whilte (1978) [5]. Further, they are also equivalent with Eqs. (4) in Koshizuka and Hayashi (1988) [1].

$E_\sigma$ and $E_\gamma$ are equal to the energies of FM magnon $\hbar\omega_{\rm FM}$ and of AFM magnon $\hbar\omega_{\rm AFM}$, respectively. In this way, $\sigma_{y,z}$ and $\gamma_{y,z}$ describe the dynamics of FM and AFM magnons, respectively.

The equations of motion of the spins in Eq. (F25) are rewritten as

$$\gamma^{-1}\dot{S}_z \approx aS_y - bS_{y'}, \tag{F29a}$$
$$\gamma^{-1}\dot{S}_{z'} \approx aS_{y'} - bS_y, \tag{F29b}$$
$$\gamma^{-1}\dot{S}_y \approx cS_z + dS_{z'}, \tag{F29c}$$
$$\gamma^{-1}\dot{S}_{y'} \approx cS_{z'} + dS_z, \tag{F29d}$$

where the coefficients are defined as

$$a = (S/g\mu_{\rm B})\left[-2z\bar{J}_{\rm Fe} - 2A_z - 2(A_x - A_z)\cos^2\beta_0\right], \tag{F30a}$$
$$b = (S/g\mu_{\rm B})(2zJ_{\rm Fe}), \tag{F30b}$$
$$c = (S/g\mu_{\rm B})\left[2z\bar{J}_{\rm Fe} + 2(A_x - A_z)\cos(2\beta_0)\right], \tag{F30c}$$
$$d = (S/g\mu_{\rm B})(-2z\bar{J}_{\rm Fe}). \tag{F30d}$$

### 3. Nonlinear dynamics of magnons in absence of external magnetic field

In order to describe the spin dynamics based on the FM and AFM magnon modes, let us rewrite Eqs. (F16) in terms of $\sigma_{x,y,z}$ and $\gamma_{x,y,z}$. Using the following relation derived from Eqs. (F26)

$$S_x = S + \frac{\gamma_x + \sigma_x}{2}, \qquad S_y = \frac{\sigma_y + \gamma_y}{2}, \qquad S_z = \frac{\sigma_z + \gamma_z}{2}, \tag{F31a}$$
$$S_{x'} = S + \frac{\gamma_x - \sigma_x}{2}, \qquad S_{y'} = \frac{\sigma_y - \gamma_y}{2}, \qquad S_{z'} = \frac{\sigma_z - \gamma_z}{2}, \tag{F31b}$$

Eqs. (F16) are written *without any approximation* as

$$\begin{aligned}\hbar\dot{\sigma}_x &= \left[-2zJ_{\rm Fe}\sin(2\beta_0) + zD_{\rm Fe}\cos(2\beta_0) - (A_x - A_z)\sin(2\beta_0)\right]\left(S + \frac{\gamma_x}{2}\right)\sigma_y \\ &+ \frac{2zJ_{\rm Fe}\sin(2\beta_0) - zD_{\rm Fe}\cos(2\beta_0) - (A_x - A_z)\sin(2\beta_0)}{2}\sigma_x\gamma_y \\ &+ \left[z(\bar{J}_{\rm Fe} - J_{\rm Fe}) + A_z\cos^2\beta_0 + A_x\sin^2\beta_0\right]\sigma_z\gamma_y \\ &+ \left[-z(\bar{J}_{\rm Fe} - J_{\rm Fe}) + A_z\cos^2\beta_0 + A_x\sin^2\beta_0\right]\sigma_y\gamma_z,\end{aligned} \tag{F32a}$$

$$\begin{aligned}\hbar\dot{\sigma}_y &= 2(A_x - A_z)\cos(2\beta_0)S\sigma_z + 2(A_x - A_z)\sin(2\beta_0)S\sigma_x \\ &+ (A_x - A_z)\cos(2\beta_0)(\sigma_z\gamma_x + \sigma_x\gamma_z) + (A_x - A_z)\sin(2\beta_0)(\sigma_x\gamma_x - \sigma_z\gamma_z),\end{aligned} \tag{F32b}$$

$$\begin{aligned}\hbar\dot{\sigma}_z &= -2\left[z(\bar{J}_{\rm Fe} + J_{\rm Fe}) + A_z + (A_x - A_z)\cos^2\beta_0\right]\left(S + \frac{\gamma_x}{2}\right)\sigma_y \\ &+ \left[z(\bar{J}_{\rm Fe} + J_{\rm Fe}) - A_z - (A_x - A_z)\cos^2\beta_0\right]\sigma_x\gamma_y \\ &+ \frac{-2zJ_{\rm Fe}\sin(2\beta_0) + zD_{\rm Fe}\cos(2\beta_0) + (A_x - A_z)\sin(2\beta_0)}{2}\sigma_z\gamma_y \\ &+ \frac{2zJ_{\rm Fe}\sin(2\beta_0) - zD_{\rm Fe}\cos(2\beta_0) + (A_x - A_z)\sin(2\beta_0)}{2}\sigma_y\gamma_z,\end{aligned} \tag{F32c}$$

$$\hbar\dot{\gamma}_x = [-2zJ_{\rm Fe}\sin(2\beta_0) + zD_{\rm Fe}\cos(2\beta_0) - (A_x - A_z)\sin(2\beta_0)]\left(S + \frac{\gamma_x}{2}\right)\gamma_y$$
$$+ \frac{2zJ_{\rm Fe}\sin(2\beta_0) - zD_{\rm Fe}\cos(2\beta_0) - (A_x - A_z)\sin(2\beta_0)}{2}\sigma_x\sigma_y$$
$$+ \left[z(\bar{J}_{\rm Fe} + J_{\rm Fe}) + A_z\cos^2\beta_0 + A_x\sin^2\beta_0\right]\sigma_z\sigma_y$$
$$+ \left[-z(\bar{J}_{\rm Fe} + J_{\rm Fe}) + A_z\cos^2\beta_0 + A_x\sin^2\beta_0\right]\gamma_z\gamma_y, \tag{F33a}$$
$$\hbar\dot{\gamma}_y = 2\left[2z\bar{J}_{\rm Fe} + (A_x - A_z)\cos(2\beta_0)\right]\left(S + \frac{\gamma_x}{2}\right)\gamma_z$$
$$+ \left[-2z\bar{J}_{\rm Fe} + (A_x - A_z)\cos(2\beta_0)\right]\sigma_z\sigma_x$$
$$+ \frac{2zJ_{\rm Fe}\sin(2\beta_0) - zD_{\rm Fe}\cos(2\beta_0) - (A_x - A_z)\sin(2\beta_0)}{2}\left({\sigma_z}^2 - {\sigma_x}^2\right), \tag{F33b}$$
$$\hbar\dot{\gamma}_z = -2\left[z(\bar{J}_{\rm Fe} - J_{\rm Fe}) + A_z + (A_x - A_z)\cos^2\beta_0\right]\left(S + \frac{\gamma_x}{2}\right)\gamma_y$$
$$+ \left[z(\bar{J}_{\rm Fe} - J_{\rm Fe}) - A_z - (A_x - A_z)\cos^2\beta_0\right]\sigma_x\sigma_y$$
$$+ \frac{-2zJ_{\rm Fe}\sin(2\beta_0) + zD_{\rm Fe}\cos(2\beta_0) + (A_x - A_z)\sin(2\beta_0)}{2}\sigma_z\sigma_y$$
$$+ \frac{2zJ_{\rm Fe}\sin(2\beta_0) - zD_{\rm Fe}\cos(2\beta_0) + (A_x - A_z)\sin(2\beta_0)}{2}\gamma_z\gamma_y, \tag{F33c}$$
$$\tag{F33d}$$

Under the assumption of $S \gg \sigma_{x,y,z}, \gamma_{x,y,z}$ and $\sigma_{z,y} \gg \sigma_x$, these equations are approximately reduced to Eqs. (F27).

Let us discuss the nonlinear dynamics of magnons in a perturbative way. Although we approximated $S_{x_i}, S_{x'_j} \approx S$ in the previous subsection, according to the above equation, they also oscillate as

$$\hbar\dot{\sigma}_x/S \approx [-2zJ_{\rm Fe}\sin(2\beta_0) + zD_{\rm Fe}\cos(2\beta_0) - (A_x - A_z)\sin(2\beta_0)]\,\sigma_y, \tag{F34a}$$
$$\hbar\dot{\gamma}_x/S \approx [-2zJ_{\rm Fe}\sin(2\beta_0) + zD_{\rm Fe}\cos(2\beta_0) - (A_x - A_z)\sin(2\beta_0)]\,\gamma_y. \tag{F34b}$$

In terms of $S_x$ and $S_{x'}$, the approximated equations of motion of them are expressed as

$$\hbar\dot{S}_x/S \approx [-2zJ_{\rm Fe}\sin(2\beta_0) + zD_{\rm Fe}\cos(2\beta_0) - (A_x - A_z)\sin(2\beta_0)]\,S_y, \tag{F35a}$$
$$\hbar\dot{S}_{x'}/S \approx [2zJ_{\rm Fe}\sin(2\beta_0) - zD_{\rm Fe}\cos(2\beta_0) + (A_x - A_z)\sin(2\beta_0)]\,S_{y'}. \tag{F35b}$$

Since $\sigma_y$ and $\gamma_y$ oscillate with $\omega_{\rm FM}$ and $\omega_{\rm AFM}$, respectively, in the limit of linear response, $\sigma_x$ and $\gamma_x$ also oscillate as

$$\sigma_x(t) \propto {\rm e}^{-{\rm i}\omega_{\rm FM}t}, \tag{F36a}$$
$$\gamma_x(t) \propto {\rm e}^{-{\rm i}\omega_{\rm AFM}t} \tag{F36b}$$

Then, let us focus on the nonlinear terms in Eqs. (F32) and (F33). Considering the situation of the present study, we here assume that only the spin dynamics $\sigma_{x,y,z}$ in the FM mode is resonantly excited, i.e., $|\gamma_{x,y,z}| \ll |\sigma_{x,y,z}|$. As seen in Eqs. (F33), through the terms proportional to $\sigma_x\sigma_y$, $\sigma_z\sigma_y$, $\sigma_z\sigma_x$, ${\sigma_x}^2$, and ${\sigma_z}^2$, the spin dynamics $\gamma_{x,y,z}$ in the AFM mode are excited with the frequency of $2\omega_{\rm FM}$, i.e., one AFM magnon is excited by two FM magnons.